\documentclass{aa}  

\usepackage{graphicx}
\usepackage{txfonts}
\usepackage{amsmath}
\usepackage{multirow,multicol}
\usepackage{enumitem}
\usepackage{color,xcolor}
\usepackage{ulem}
\usepackage{makecell,booktabs,threeparttable}
\usepackage{mhchem}
\usepackage{soul}

\newcommand\hh{\ce{H2}}
\newcommand\hho{\ce{H2O}}
\newcommand\hooh{\ce{H2O2}}
\newcommand\oo{\ce{O2}}
\newcommand\ooo{\ce{O3}}
\newcommand\ooh{\ce{O2H}}

\newcommand\hoohbyhho{[\ce{H2O2}]/[\ce{H2O}]}
\newcommand\hhbyhho{[\ce{H2}]/[\ce{H2O}]}
\newcommand\oobyhho{[\ce{O2}]/[\ce{H2O}]}
\newcommand\ooobyoo{[\ce{O3}]/$[\ce{O2}]_0$}

\usepackage[hidelinks,colorlinks=true,linkcolor=blue,citecolor=blue]{hyperref}

\begin{document}

   \title{Simulation of proton radiolysis of \hho\ and \oo\ ices with the Nautilus code}

   \author{Tian-Yu Tu\inst{1,2}
          \and
          Valentine Wakelam\inst{2}
          \and
          Jean-Christophe Loison\inst{3}
          \and
          Marin Chabot\inst{4}
          \and
          Emmanuel Dartois\inst{5}
          \and
          Yang Chen\inst{1,6}
          }

   \institute{
             School of Astronomy \& Space Science, Nanjing University, 163 Xianlin Avenue, Nanjing 210023, China\\
             \email{ygchen@nju.edu.cn}
         \and
             Laboratoire d'Astrophysique de Bordeaux, Univ. Bordeaux, CNRS, B18N, allée Geoffroy Saint-Hilaire, 33615 Pessac, France
             \email{valentine.wakelam@u-bordeaux.fr}
         \and
             Institut des Sciences Moléculaires (ISM), CNRS, Univ. Bordeaux, 351 cours de la Libération, 33400 Talence, France
         \and 
             Laboratoire des deux infinis Irène Joliot Curie (IJClab), CNRS-IN2P3, Université Paris-Saclay, 91405 Orsay, France
         \and
             Institut des Sciences Moléculaires d’Orsay (ISMO), UMR8214, CNRS, Université Paris-Saclay, Bât 520, Rue André Rivière,
          91405 Orsay, France
         \and
             Key Laboratory of Modern Astronomy and Astrophysics, Nanjing University, Ministry of Education, Nanjing 210023, China
             }

   \date{Received ...; accepted ...}

 
  \abstract
   {The radiolysis effect of cosmic rays (CRs) play an important role in the chemistry in molecular clouds. 
   CRs can dissociate the molecules on dust grains, producing reactive suprathermal species and radicals which facilitate the formation of large molecules. 
   }
   {In this study, we try to reproduce the laboratory experimental results on radiolysis of pure \hho\ and \oo\ ice with the Nautilus astrochemical code, and evaluate the effects of changing some of the uncertain physical and chemical parameters. 
   }
  {We add the radiolysis reactions, quenching reactions of suprathermal species, and reactions between suprathermal and thermal species into the Nautilus code. 
  By adjusting the parameters in the code, we investigate the sensitivity of the simulation results of the \hho\ ice on the removal of reaction-diffusion competition, the removal of non-diffusive chemistry, and the desorption energies of the suprathermal species.}
   {We find that the Nautilus model, with a few adjustments of the chemistry, can reproduce the steady-state \hoohbyhho\ and \ooobyoo\ abundance ratios in the \hho\ and \oo\ radiolysis experiments at any CR flux in the experiments. 
   These adjustments in the model do not fully reproduce the fluence required to reach the steady state. It tends also to overestimate the destruction of \hho\ as measured in \hho\ radiolysis experiments. We show that reducing the $G$-values of \hho\ radiolysis, which implies an increase in the efficiency of immediate reformation of water molecules locally after ion impact, leads to simulated \hho\ destruction rates closer to the experiments.
   The effect of reaction-diffusion competition on the simulation results of \hho\ ice is significant at $\zeta \lesssim 10^{-14}\rm \ s^{-1}$ only. 
   The non-diffusive chemistry affects the simulation results at 16 K but not 77K, while the results are sensitive to the desorption energies of suprathermal H, O, $\ce{O3}$ and OH at 77 K. 
   }
   {Our results show that the steady-state \hoohbyhho\ and \ooobyoo\ in radiolysis experiments can be reproduced by fine-tuning the chemical model, but still call for more constraints on the intermediate pathways in the radiolysis processes, especially the ion chemistry in the ice bulk, as well as activation barriers and branching ratios of the reactions in the network.}


   \maketitle
%

\section{Introduction} \label{sec:intro}
Cosmic rays (CRs) are energetic ($\rm \sim 10^6$--$10^{18}$ eV) charged particles that consist of protons, heavy ions and electrons. 
They permeate our Galaxy and strongly influence the physics and chemistry of the interstellar medium \citep[ISM, e.g.,][]{Dalgarno_Interstellar_2006,Indriolo_Cosmic-ray_2013,Gaches_High-energy_2025}. 
The heating effect of CRs plays an important role in the thermal balance of the ISM \citep{Goldsmith_Molecular_1978}. 
Low-energy CRs act as the primary source of ionization in dense molecular clouds (MCs) shielded from UV radiation \citep{Padovani_Cosmic-ray_2009,Chabot_Cosmic-ray_2016}. 
CRs can ionize the molecular hydrogen through the reaction
\begin{equation}
    \rm H_2 \xrightarrow[]{CR} H_2^+ + e^-, 
\end{equation}
whose rate coefficient is defined as the CR ionization rate per \hh\ molecule which is widely used to characterize the degree of CR ionization in MCs, typically of the order $10^{-17}$--$10^{-16}\rm \ s^{-1}$ \citep[e.g.,][]{Caselli_Ionization_1998,Indriolo_Investigating_2012,Neufeld_Cosmic-Ray_2017,Bovino_new_2020,Neufeld_Densities_2024a,Obolentseva_Reevaluation_2024,Bialy_first_2025,Neufeld_JWST_2025,Indriolo_Mapping_2026}. 
$\rm H_2^+$ then reacts with H$_2$ to form H$_3^+$, which initiates the formation of poly-atomic molecules in the gas-phase in MCs. 
CRs can also ionize He to form $\rm He^+$ which can in turn dissociate CO in MCs to form C and $\rm C^+$, which affects our understanding of the molecular inventory in galaxies \citep{Bisbas_Effective_2015,Bisbas_Cosmic-ray_2017}.
In addition, CRs are believed to be a source of UV radiation inside dense MCs \citep{Prasad_UV_1983}. 

\par

In addition to ionization in the gas-phase, CRs can release the molecular species on the dust mantle back to the gas phase via global heating of grains \citep{Leger_Desorption_1985,Hasegawa_New_1993}, impulsive spot heating of grains \citep{Ivlev_Impulsive_2015}, photo-desorption by secondary UV photons induced by CRs \citep{Oberg_Photodesorption_2007,Ruaud_Gas_2016}, sputtering \citep{Dartois_Cosmic_2018,Wakelam_Efficiency_2021}, etc. 
These non-thermal desorption processes have been proposed to be at the origin of complex organic molecules (COMs) observed in cold cores \citep[e.g.,][]{Bacmann_Detection_2012}. 
Radiolysis is another important process of CRs that can facilitate formation of COMs in cold cores \citep[e.g.,][]{Pilling_Radiolysis_2010,deBarros_Radiolysis_2011}. 
When CRs collide with dust grains, the primary CR particles and secondary electrons can dissociate the molecules in the ice mantles to form radicals and suprathermal species, which are highly reactive and can enhance the production of COMs \citep{Shingledecker_Cosmic-Ray-driven_2018}. 
This effect was extensively studied by laboratory experiments.
For example, HCOOH and $\rm CH_3OH$ are produced after the irradiation of a mixture of \hho\ and $\rm CO_2$ ice by 0.8 MeV protons \citep{Hudson_Laboratory_1999}. 
Similar experiments have also been conducted with energetic electrons \citep[e.g.,][]{Bennett_Laboratory_2005a,Zheng_Temperature_2006} and heavy ions as CR analogues \citep[e.g.,][]{Palumbo_ROCN_2000,Baratta_comparison_2002,SeperueloDuarte_Laboratory_2010,Boduch_Chemistry_2012}. 
These experiments strongly suggest that radiolysis can alter the chemical composition of ice and trigger the formation of large molecules. 

\par

Radiolysis needs to be considered in complex astrochemical models in order to be more realistic. Before doing so, the model needs to be benchmarked over experiments. 
\citet{Shingledecker_general_2018} proposed a method to introduce ice radiolysis into chemical models based on rate equations. 
They explored the results of radiolysis in a dark MC with a complex chemical model. 
By incorporating radiolysis into the MONACO code \citep{Vasyunin_Unified_2013,Vasyunin_Formation_2017}, \citet[][hereafter SVHC19 for short]{Shingledecker_Simulating_2019} successfully reproduced the experimental results of \hho\ and \oo\ ice irradiated by protons \citep{Baragiola_Solid-state_1999,Gomis_Hydrogen_2004}. 
Using the same method, the simulation of \citet{Paulive_role_2021} suggested that the abundances of some COMs, such as $\rm HCOOCH_3$ and $\rm CH_2OHCHO$, in dark MCs can be enhanced by the radiolysis process. 
Similar method was also used to simulate the experiment of \oo\ ice exposed to ionizing UV radiation \citep{Mullikin_New_2021}. 
On the other hand, \citet{Pilling_Chemical_2022,Pilling_Chemical_2023} simulated the radiolysis experiments of \hho\ and $\ce{CO2}$ ices, and obtain the reaction rates by fitting to the experimental results. 

\par

In this paper, we expand the work of SVHC19 on simulating the proton-irradiated \hho\ and \oo\ ice using another chemical model, Nautilus. 
In a sophisticated chemical model, many parameters are tunable with often scarce constraints from experiments or theoretical calculations. 
We explore the influence of some parameters on the simulated chemistry that occurs in irradiated \hho\ and \oo\ ice. 
The paper is organized as follows.
In Section \ref{sec:model&network} we introduce how we modify the chemical model and construct the reaction network. 
The simulation results of \hho\ and \oo\ and several test models are shown in Section \ref{sec:res}, and are discussed in Section \ref{sec:disc}. 
We summarize our findings in Section \ref{sec:con}.

\section{Model and Network} \label{sec:model&network}
\subsection{Chemical model} \label{sec:model}

The simulation is based on the \texttt{Nautilus} code, which is a three-phase rate-equation chemical model \citep{Ruaud_Gas_2016}. 
The experiments that we will consider are done in thick ices with little to no interactions with the gas. So to simulate the chemical processes in the experiments, we consider only bulk species and remove all the gas-phase reactions, surface-bulk swapping, adsorption and desorption, and photodissociation by CR-induced UV photons. 
Following SVHC19, we add the following processes of radiolysis to the code: 
\begin{align}
\rm A  &\rm \leadsto A^+ + e^- \rightarrow A^* \rightarrow B^* + C^*, \tag{R1} \label{reac:radiolysis1}\\
\rm A  &\rm \leadsto A^* \rightarrow B + C, \tag{R2} \label{reac:radiolysis2}\\ 
\rm A  &\rm \leadsto A^*, \tag{R3}\label{reac:radiolysis3}
\end{align}
where the curly arrow means the dissociation reaction due to collision with CR protons or secondary electrons, and the species with asterisk are suprathermal species. 
In process \ref{reac:radiolysis1}, the target species A is ionized by the incident particle, but soon recombine to form suprathermal $\rm A^*$ which is then dissociated into suprathermal $\rm B^*$ and $\rm C^*$. 
In processes \ref{reac:radiolysis2} and \ref{reac:radiolysis3}, the species A is directly excited to suprathermal state, and in process \ref{reac:radiolysis2} the suprathermal $\rm A^*$ is dissociated into $\rm B^*$ and $\rm C^*$. 
The reaction rate coefficients of these processes can be approximated in first order to \citep{Shingledecker_general_2018}: 
\begin{equation}
\label{eq:k_radiolysis}
    k=S_{\rm e}\left(\frac{G}{100\,\rm eV}\right)  \phi\left(\frac{\zeta}{10^{-17}\,\rm s^{-1}}\right), 
\end{equation}
where $S_{\rm e}$ is the electronic stopping power of species A at a given energy of the incident particle, the $G$-value is the number of molecules destroyed after every unit energy of 100 eV is transferred from the incident particles to the irradiated ice \citep{Dewhurst_General_1952}, $\phi=8.6\rm \, particles\, cm^{-2}\, s^{-1}$ is the integrated interstellar CR flux \citep{Spitzer_Heating_1968}, and $\zeta$ is the CR ionization rate. 
All of the simulated radiolysis processes and their $G$ values are adopted from \citet{Shingledecker_Cosmic-Ray-driven_2018} and SVHC19, and are listed in Table \ref{tab:radiolysis_reactions}. 
Rather than following SVHC19 to use the same stopping power $S_{\rm e}$ of \hho\ from \citet{Baragiola_Solid-state_1999} for all the radiolysis reactions of different species, we computed them for each molecular species with the SRIM code \citep{Ziegler_SRIM_2010}. In Table \ref{tab:radiolysis_reactions}, we also list the $S_{\rm e}$ values  for two different energies, which are the energies of impacting particles used in the experiments we will use as references.

\begin{table}
\caption{Radiolysis processes considered in the simulation with their branching ratios, $G$-values, and radiolysis process type, as well as the calculated stopping powers of the considered species in units of $10^{-14}\ \rm eV\, cm^2\, molecule^{-1}$.}
\label{tab:radiolysis_reactions}
\centering
\begin{tabular}{lccc}
\hline
\hline
Reaction & $f_{\rm br}$ & $G$-value & Type \\
\hline
\multicolumn{4}{c}{\ce{H2O}} \\
\multicolumn{4}{c}{$S_{\rm e}(200\ {\rm keV})=1.93$} \\
$\mathrm{H_2O \leadsto OH^* + H^*}$ & 0.9 & 3.704 & R1 \\
$\mathrm{H_2O \leadsto O^* + H_2^*}$ & 0.1 & 3.704 & R1 \\
$\mathrm{H_2O \leadsto OH + H}$ & 0.9 & 1.747 & R2 \\
$\mathrm{H_2O \leadsto O + H_2}$ & 0.1 & 1.747 & R2 \\
$\mathrm{H_2O \leadsto H_2O^*}$ & 1.0 & 1.747 & R3 \\ \hline
\multicolumn{4}{c}{\ce{O2}} \\
\multicolumn{4}{c}{$S_{\rm e}(5\ {\rm keV})=0.93$, $S_{\rm e}(200\ {\rm keV})=2.61$} \\
$\mathrm{O_2 \leadsto O^* + O^*}$ & 1.0 & 3.704 & R1 \\
$\mathrm{O_2 \leadsto O + O}$ & 1.0 & 2.138 & R2 \\
$\mathrm{O_2 \leadsto O_2^*}$ & 1.0 & 2.138 & R3 \\ \hline
\multicolumn{4}{c}{\ce{O3}} \\
\multicolumn{4}{c}{$S_{\rm e}(5\ {\rm keV})=1.42$, $S_{\rm e}(200\ {\rm keV})=4.00$} \\ 
$\mathrm{O_3 \leadsto O_2^* + O^*}$ & 1.0 & 3.704 & R1 \\
$\mathrm{O_3 \leadsto O_2 + O}$ & 1.0 & 4.059 & R2 \\
$\mathrm{O_3 \leadsto O_3^*}$ & 1.0 & 4.059 & R3 \\ \hline
\multicolumn{4}{c}{\ooh} \\
\multicolumn{4}{c}{$S_{\rm e}(200\ {\rm keV})=3.02$} \\
$\mathrm{O_2 \leadsto O^* + OH^*}$ & 0.5 & 3.704 & R1 \\
$\mathrm{O_2H \leadsto H^* + O_2^*}$ & 0.5 & 3.704 & R1 \\
$\mathrm{O_2H \leadsto O + OH}$ & 0.5 & 3.714 & R2 \\
$\mathrm{O_2H \leadsto H + O_2}$ & 0.5 & 3.714 & R2 \\
$\mathrm{O_2H \leadsto HO_2^*}$ & 1.0 & 3.714 & R3 \\ \hline
\multicolumn{4}{c}{\hooh} \\
\multicolumn{4}{c}{$S_{\rm e}(200\ {\rm keV})=3.38$} \\
$\mathrm{H_2O_2 \leadsto OH^* + OH^*}$ & 0.5 & 3.704 & R1 \\
$\mathrm{H_2O_2 \leadsto H^* + O_2H^*}$ & 0.5 & 3.704 & R1 \\
$\mathrm{H_2O_2 \leadsto OH + OH}$ & 0.5 & 2.296 & R2 \\
$\mathrm{H_2O_2 \leadsto H + O_2H}$ & 0.5 & 2.296 & R2 \\
$\mathrm{H_2O_2 \leadsto H_2O_2^*}$ & 1.0 & 2.296 & R3 \\ \hline
\multicolumn{4}{c}{\ce{O}} \\
\multicolumn{4}{c}{$S_{\rm e}(5\ {\rm keV})=0.47$, $S_{\rm e}(200\ {\rm keV})=1.33$} \\ 
$\mathrm{O \leadsto O^*}$ & 1.0 & 3.70 & R1 \\
$\mathrm{O \leadsto O^*}$ & 1.0 & 1.93 & R3 \\ \hline
\multicolumn{4}{c}{\ce{H2}} \\
\multicolumn{4}{c}{$S_{\rm e}(200\ {\rm keV})=0.35$} \\
$\mathrm{H_2 \leadsto H^* + H^*}$ & 1.0 & 3.70 & R1 \\
$\mathrm{H_2 \leadsto H + H}$ & 1.0 & 1.02 & R2 \\
$\mathrm{H_2 \leadsto H_2^*}$ & 1.0 & 1.02 & R3 \\ \hline
\multicolumn{4}{c}{\ce{OH}} \\
\multicolumn{4}{c}{$S_{\rm e}(200\ {\rm keV})=1.69$} \\
$\mathrm{OH \leadsto O^* + H^*}$ & 1.0 & 3.70 & R1 \\
$\mathrm{OH \leadsto O + H}$ & 1.0 & 5.66 & R2 \\
$\mathrm{OH \leadsto OH^*}$ & 1.0 & 5.66 & R3 \\
\hline
\end{tabular}
\end{table}

Suprathermal species either are quenched by the bulk mantle to thermal species, or react both diffusively and non-diffusively with another thermal icy species. 
The quenching reaction can be written as: 
\begin{equation}
    \rm A^*  \longrightarrow A , \tag{R4}
\end{equation}
where A is an arbitrary species. 
Its rate coefficient (in units of $\rm s^{-1}$) is the characteristic vibrational frequency of species A: 
\begin{equation}
    \nu_{\rm A} = \sqrt{\frac{2N_{\rm site} E_{\rm D}^{\rm A}}{\pi^2m_{\rm A}}}, 
\end{equation}
where $E_{\rm D}^{\rm A}$ (in K) and $m_{\rm A}$ (in g) are the desorption energy and mass of species A, and $N_{\rm site}$ is the number of sites on each grain. 

\par

Reaction between suprathermal and thermal species (denoted as $\rm A^*$ and B, respectively) can proceed in both diffusive and non-diffusive manner. 
The rate coefficient of diffusive reaction was introduced in \citet{Ruaud_Gas_2016}: 
\begin{equation}
\label{eq:k_diff}
    k_{\rm diff}=f_{\rm br}\kappa_{\rm AB}\left[ \nu_{\rm A}\exp{\left(-\frac{E_{\rm diff}^{\rm A}}{T_{\rm ice}}  \right)} + \nu_{\rm B}\exp{\left(-\frac{E_{\rm diff}^{\rm B}}{T_{\rm ice}}  \right)}\right]\frac{1}{N_{\rm site}n_{\rm dust}}, 
\end{equation}
where $f_{\rm br}$ is the branching ratio, $\kappa_{\rm AB}$ is the probability for the reaction to occur, $E_{\rm diff}^{\rm A}$ and $E_{\rm diff}^{\rm B}$ are the diffusion barriers for A and B, respectively (in K), and $n_{\rm dust}$ is the gas-phase concentration of dust particles (in cm$^{-3}$). 
We assume $\kappa_{\rm AB}=1$ for exothermic and barrierless reactions \citep{Hasegawa_Models_1992}, but calculate $\kappa_{\rm AB}$ as the result of the competition between reaction and diffusion for exothermic reactions with activation barriers \citep{Chang_Gas-grain_2007,Garrod_Formation_2011,Ruaud_Gas_2016}. 
This modification of the surface rate coefficients is done to mimic the fact that two species on the same site have a probability to stay longer than to diffuse away, increasing the probability of the reaction to occur despite the barrier. 
Note that this competition was not included by SVHC19 for the ice bulk. 
We include this competition in our simulation, and its effect will be presented in Section \ref{sec:res_compe}. 
SVHC19 used the MONACO code that was described in \citet{Vasyunin_Formation_2017} and \citet{Vasyunin_Unified_2013}. 
\citet{Vasyunin_Unified_2013} justify this on the fact that most if not all the activation barriers for surface processes are based on gas-phase experiments where this process does not occur.
Another strong difference between Nautilus and MONACO is that the latter one includes the modified rate equations capable of taking into account stochastic effects in surface chemistry as proposed by \citet{Garrod_new_2008}.
In the models where we do not consider the competition reaction-diffusion, $\kappa_{\rm AB}$ is computed following \citet{Hasegawa_Models_1992}, i.e. assuming quantum mechanical probability for tunneling.
We will explore the influence of reaction-diffusion competition in Section \ref{sec:res_compe}. 

\par

Another important process is the non-diffusive chemistry. 
This process mimics the fact that when species are abundant, the probability to be formed closed by, by photo-processes or radiolysis for instance, is so high that they react without having to diffuse \citep{Jin_Formation_2020}. 
SVHC19 showed that taking this process into account is crucial. 
We then have included the formalism from SVHC19 with a slight modification: 
\begin{equation}
\label{eq:k_nondiff}
    k_{\text{non-diff}}= f_{\rm br}\left( \frac{\nu_{\rm A}+\nu_{\rm B}}{\max{(N_{\rm bulk},N_{\rm site})}} \right) \exp{\left(-\frac{E_{\rm act}^{\rm AB}}{T_{\rm ice}}\right)},
\end{equation}
where $N_{\rm bulk}$ is the total number of bulk species in the simulated ice instead of the number of sites on one grain as done by SVHC19. 
This modification is done to take into account the dilution of the species in the mantle in a more correct way and decreases the efficiency of non-diffusive chemistry. 
$E_{\rm act}^{\rm AB}$ is the activation barrier of the reaction. 
The formula is simply Equation~\ref{eq:k_diff} assuming that the diffusion is immediate, i.e. $\exp{( -E_{\rm diff}^{\rm A}/T_{\rm ice} ) } = 1$. 
We also included in Nautilus the formalism from \citet{Jin_Formation_2020}, which is completely different. However, with the reduced network used in this work, the code did not manage to converge as the destruction rates of minor species with the non-diffusive chemistry are much higher than their production rate. We conclude that the formalism from \citet{Jin_Formation_2020} cannot be used in this case. 

\par

SVHC19 mostly tested the effect of the efficiency of diffusion of the species in the ice by comparing three models: the first including the non-diffusive chemistry, the second including only the thermal hopping of the species, and the third in which H, \hh\ and O are allowed to diffuse through tunneling effects. 
In our case, all species are allowed to diffuse through tunneling effect but with an efficiency that depends on their mass \citep{Wakelam_2024_2024a}. 
We will explore the effect of the non-diffusive chemistry in sections \ref{sec:res_non-diff}. 

\par

The desorption energies ($E_{\rm D}$) of the simulated species are adopted from SVHC19 except for H and \hh, and are listed in Table \ref{tab:ED} in order to be able to be compared in a more quantitative way with their results. 
For H and \hh, the desorption energies are 650 K and 440 K, respectively \citep{Wakelam_Binding_2017}.
For the other species, the diffusion barriers are assumed to be $0.7E_{\rm D}$ (same as SVHC19). 
The desorption energies and diffusion barriers of suprathermal species are assumed to be the same as those of the corresponding thermal species. 
The influence of changing the desorption energies will be investigated in Section \ref{sec:res_ED}.

\begin{table}
\caption{Desorption energies of the simulated species}
\label{tab:ED}
\centering
\begin{tabular}{cc}
\hline
\hline
Species & $E_{\rm D}$ (K) \\
\hline
H & 650 \\ 
\hh & 440 \\
O & 1660 \\ 
\oo & 930 \\ 
\ooo & 1833 \\ 
OH & 2850 \\ 
\ooh & 4510 \\ 
\hho & 5700 \\ 
\hooh & 5700 \\ 
\hline
\end{tabular}
\end{table}

Throughout the simulation, we keep the gas density of $n_{\rm H}=10^9\rm \ cm^{-3}$ with a gas-to-dust mass ratio of 100. 
The change in gas density will not give rise to any changes in our simulation results. 
The CR ionization rate is set to be $1.3\times 10^{-7}\rm \ s^{-1}$, corresponding to an ion flux of $\sim 1.1 \times 10^{11}\rm\ particles\ cm^{-2}\ s^{-1}$.
This is significantly higher than the CR ionization rates in the ISM \citep{Caselli_Ionization_1998,Indriolo_Investigating_2012}, but is consistent with the flux used in the simulation of SVHC19 and is also the typical ion flux used in experiments \citep[e.g.,][]{Gomis_Hydrogen_2004a,Gomis_Hydrogen_2004}. 
We will discuss how the simulation results are affected by the values of CR ionization rates in Section \ref{sec:res_zeta}.

\subsection{Reaction network} \label{sec:network}
The considered network is shown in Table \ref{tab:network1}, which is based on Table 4 of SVHC19 for \hooh\ with some modifications. It contains 38 reactions for 9 bulk species. As already stated, we ignore gas-phase and surface reactions as well as any exchanges between them and the bulk.
For some of the reactions in this network, the activation barriers are not recorded in references. 
Following SVHC19 for the unknown barriers, we assume that the barriers are 10000 K except for radical-radical reactions, whose barriers are assumed to be 0. 
For each thermal reaction $\rm A+B \rightarrow product$ in Table \ref{tab:network1}, we include both the diffusive pathways and non-diffusive pathways for reactions involving thermal radicals and atomic O. 
The original references for the activation energies and branching ratios are given in SVHC19. Some of the values have been changed and we give the new reference in the Table. 
Reactions involving thermal atomic H do not have the non-diffusive pathways by default because H can diffuse quickly in the ice.  
We also include the diffusive and non-diffusive pathways for reactions involving suprathermal species: 
\begin{align}
    \rm A^* + B & \rm \longrightarrow product, \tag{R5} \\ 
    \rm A + B^* & \rm \longrightarrow product. \tag{R6} 
\end{align}
The activation barriers of these reactions involving suprathermal species are assumed to be of 0. 

\begin{table}
\caption{Considered reaction network based on Table 4 of SVHC19. }
\label{tab:network1}
\centering
\begin{tabular}{lcc}
\hline
\hline
Reaction & $E_{\rm A}$ (K)$^a$ & $f_{\rm br}$ \\
\hline
\ce{H + H -> H2} & 0 & 1.0  \\
\ce{H + O -> OH} & 0 & 1.0  \\
\ce{H + O2 -> O2H} & 0 & 1.0  \\
\ce{H + O3 -> O2 + OH} & 450 & 1.0 \\
\ce{H + OH -> H2O} & 0 & 1.0  \\
\ce{H + H2O -> OH + H2} & 9700 & 1.0  \\
\ce{H + O2H -> OH + OH} & 0 & 0.2$^b$  \\
\ce{H + O2H -> O2 + H2} & 0 & 0.2$^b$  \\
\ce{H + O2H -> H2O2} & 0 & 0.6$^b$  \\
\ce{H + H2O2 -> H2O + OH} & 3400$^c$ & 0.5$^c$  \\ 
\ce{H + H2O2 -> O2H + H2} & 4200$^c$ & 0.5$^c$  \\
\ce{H2 + O -> H2O} & 9700 &  1.0   \\
\ce{H2 + O2 -> O2H + H} & 28000 & 1.0  \\
\ce{H2 + OH -> H2O + H} & 1800 & 1.0  \\
\ce{H2 + O2H -> H2O2 + H} & 13000 & 1.0  \\
\ce{H2 + H2O2 -> H2 + OH + OH} & 10000 & 1.0  \\
\ce{O + O2 -> O3} & 300$^d$ & 1.0  \\
\ce{O + OH -> O2H} & 0 & 1.0  \\
\ce{O + H2O -> H2O2} & 8800 & 1.0  \\
\ce{O + O2H -> O2 + OH} & 0 & 1.0  \\
\ce{O + H2O2 -> OH + O2H} & 2000 & 1.0  \\
\ce{O2 + OH -> O2H + O} & 25000 & 1.0  \\
\ce{O2 + H2O -> O2H + OH} & 37000 & 1.0  \\
\ce{O2 + O2H -> O3 + OH} & 10000 & 1.0  \\
\ce{O2 + H2O2 -> O2H + O2H} & 18000 & 1.0  \\
\ce{O3 + OH -> O2H + O2} & 940 & 1.0  \\
\ce{O3 + H2O -> H2O2 + O2} & 10000 & 1.0  \\
\ce{O3 + O2H -> O2 + O2 + OH} & 490 & 1.0  \\
\ce{O3 + H2O2 -> O2 + O2 + H2O} & 10000 & 1.0  \\
\ce{OH + OH -> H2O2} & 0 & 1.0  \\
\ce{OH + H2O -> H2O2 + H} & 40000 & 0.5  \\ 
\ce{OH + H2O -> H2O + OH} & 2100 & 0.5  \\
\ce{OH + O2H -> H2O + O2} & 0 & 1.0  \\
\ce{OH + H2O2 -> H2O + O2H} & 220$^e$ & 1.0  \\
\ce{H2O + O2H -> H2O2 + OH} & 17000 & 1.0  \\
\ce{H2O + H2O2 -> OH + OH + H2O} & 10000 & 1.0 \\
\ce{O2H + O2H -> H2O2 + O2} & 0 & 0.9$^f$  \\
\ce{O2H + O2H -> H2O + O3} & 0 & 0.1$^f$  \\
\ce{H2O2 + H2O2 -> H2O + OH + O2H} & 10000 & 1.0 \\
\hline
\end{tabular}
\begin{tablenotes}
\footnotesize 
\item $a$: Reactions with $E_{\rm A}=10000$ K are assumed because of lack of experimental values (SVHC19). 
For radical-radical reactions without experimental values, the activation barriers are assumed to be 0. 
\item $b$: \citet{Cuppen_Water_2010,Lamberts_Water_2013}
\item $c$: \citet{Ellingson_Reactions_2007,Lamberts_Water_2013,Lamberts_Quantum_2016,Lamberts_Influence_2017,Lu_Rate_2018,Lu_Theoretical_2019}
\item $d$: \citet{Benderskii_Diffusion-limited_1996} 
\item $e$: \citet{Atkinson_Evaluated_2004,Buszek_Effects_2012} 
\item $f$: Formation of $\ce{H2O2 + O2}$ is the major pathway of the $\ce{O2H + O2H}$ reaction, while the formation of $\ce{H2O + O3}$ is a minor pathway \citep{Zhang_Water-catalyzed_2011,Silva_Molecular_2025}. Here we assume branching ratios of 0.9 and 0.1 for these two pathways, respectively. 
\end{tablenotes}
\end{table}

In addition to the reactions in Table \ref{tab:network1}, we also consider the reactions of O, \oo, and \ooo\ in Table 7 of \citet{Shingledecker_new_2017}, and this part of the network is listed in Table \ref{tab:network2}. 
The electronically excited species in \citet{Shingledecker_new_2017} are treated here as suprathermal species which can react with thermal species without barrier. 
We include both diffusive and non-diffusive pathways for reactions in Table \ref{tab:network2}.

\begin{table}
\caption{Considered reaction network based on Table 7 of \citet{Shingledecker_new_2017}. }
\label{tab:network2}
\centering
\begin{tabular}{lcc}
\hline
\hline
Reaction & $E_{\rm A}$ (K) & $f_{\rm br}$ \\
\hline
$\mathrm{O} + \mathrm{O} \longrightarrow \mathrm{O_2^*}$ & 0 & 1.0  \\
$\mathrm{O} + \mathrm{O_2} \longrightarrow \mathrm{O_3^*}$ & 300$^a$ & 1.0  \\
$\mathrm{O} + \mathrm{O_3} \longrightarrow 2\mathrm{O}_2$  & 2060 & 1.0 \\
$\mathrm{O_3} + \mathrm{O_3} \longrightarrow 3\mathrm{O_2}$ & 9310 & 1.0  \\
$\mathrm{O^*} + \mathrm{O} \longrightarrow 2\mathrm{O}$ & 0 & 0.5   \\
$\mathrm{O^*} + \mathrm{O} \longrightarrow  \mathrm{O_2^*}$ & 0 & 0.5  \\
$\mathrm{O^*} + \mathrm{O_2} \longrightarrow \mathrm{O} + \mathrm{O_2}$ & 0 & 0.9$^a$  \\
$\mathrm{O^*} + \mathrm{O_2} \longrightarrow \mathrm{O_3^*}$ & 0 & 0.05$^a$  \\
$\mathrm{O^*} + \mathrm{O_3} \longrightarrow 2\mathrm{O}_2$  & 0 & 1.0  \\
$\mathrm{O_2^*} + \mathrm{O} \longrightarrow \mathrm{O} + \mathrm{O_2}$ & 0 & 0.333$^b$  \\
$\mathrm{O_2^*} + \mathrm{O} \longrightarrow \mathrm{O_3^*}$ & 0 & 0.333$^b$  \\
$\mathrm{O_2^*} + \mathrm{O_2} \longrightarrow 2\mathrm{O_2}$ & 0 & 1.0  \\
$\mathrm{O_2^*} + \mathrm{O_3} \longrightarrow 2\mathrm{O_2} + \mathrm{O}$ & 0 & 1.0  \\
$\mathrm{O_3^*} + \mathrm{O} \longrightarrow 2\mathrm{O}_2$  & 0 & 1.0  \\
$\mathrm{O_3^*} + \mathrm{O_2} \longrightarrow 2\mathrm{O}_2 + \mathrm{O}$  & 0 & 1.0 \\
$\mathrm{O_3^*} + \mathrm{O_3} \longrightarrow 3\mathrm{O_2}$ & 0 & 1.0 \\
\hline
\end{tabular}
\begin{tablenotes}
\footnotesize 
\item $a$: \citet{Benderskii_Diffusion-limited_1996}
\item $b$: The branching ratios of the reactions with $\mathrm O^*$ and $\mathrm O_2$ are set to be 0.05, 0.9 and 0.05 when they produces $\mathrm O_3$ (in the suprathermal version of Table \ref{tab:network1}), $\mathrm O+ \mathrm O_2$ and $\mathrm O_3^*$, respectively. 
\item $c$: The branching ratios of these two reactions are 0.333 because there are reactions in the suprathermal version of Table \ref{tab:network1} that share the same reactants with them. 
\end{tablenotes}
\end{table}

\subsection{Conversion among ion fluence, energy dose, and simulation time}

The abundances of the simulated species in the chemical model are given as a function of simulation time ($t$). 
However, the abundances of species in the experiments are given as a function of ion fluence or energy dose, with 
\begin{equation}
    \text{dose}=\text{fluence} \times S_{\rm e}. 
\end{equation}
The ion fluence is defined as the total number of impinging proton per area in units of $\rm protons\,cm^{-2}$ and can be directly monitored in experiments, while the energy dose is the total energy absorbed by the target species in units of $\rm eV\,molecule^{-1}$.
As a first-order approximation, the dose is the governing parameter that determine the result of the radiolysis process.
The relation between the experimental dose and the simulated time is calculated as follow. 
We first adopt an approximated equation for the CR spectral density \citep{Shen_Cosmic_2004}: 
\begin{equation}
    j_n=\frac{9.42\times 10^4 E^{0.3}}{(E+E_0)^3}\rm \, particles\,cm^{-2}\,s^{-1}\,sr^{-1}\,(MeV/nucl)^{-1}, 
    \label{eq:CRden}
\end{equation}
where $E$ is the CR energy and $E_0$ is a form parameter between 0 and 940 MeV which determines the spectra of low-energy CRs. 
To determine the value of $E_0$, we fit the normalization \citep{Padovani_Cosmic-ray_2009}: 
\begin{equation}
    1.7\times 4\pi \sum_Z f_Z \int^{10\,\rm GeV}_{0.1\,\rm MeV}j_n(E,Z) \sigma(E,Z)\,dE = 10^{-17}\rm \, s^{-1}
    \label{eq:CRcal}
\end{equation}
where $Z$ is the nuclear charge of the CR species, $f_Z$ is the abundance of the CR species, $f_Z$ is the abundance of the CR species $Z$, $j_n(E,Z)$ is the CR spectral density of a CR species $Z$ with an energy of $E$, $\sigma(E,Z)=Z^2\sigma(E,Z=1)$ is the $\rm H_2$ ionization cross section of a CR species $Z$ with an energy of $E$ adopted from \citet{Rudd_Cross_1983}, and $10^{-17}\rm \ s^{-1}$ on the right hand side of the equation is the characteristic CR ionization rate will be used in Equation \ref{eq:dose}. 
The coefficient 1.7 is to consider the effect of secondary electrons during the ionization. 
We consider H, He, C, N, O, Mg, Si, and Fe in the CR constituents and adopt the abundances in \citet{Kalvans_Temperature_2018}, and obtain $E_0=637$ MeV. 
Then the relation between dose and time can be calculated by 
\begin{equation}
\label{eq:dose}
    {\rm dose} = 4\pi t\ \left(\frac{\zeta}{10^{-17}\, {\rm s^{-1}}}\right)\ \sum_Z f_Z \int^{10\,\rm GeV}_{0.1\,\rm MeV}j_n(E,Z) S_{\rm e}(E,Z)\,dE,
\end{equation}
where $\zeta$ is the CR ionization rate per $\rm H_2$ adopted in the simulation. 
After substituting the CR spectral densities of different CR constituents, $j_n(E,Z)$, obtained from Equations \ref{eq:CRden} and \ref{eq:CRcal}, as well as the stopping powers of the CR constituents, $S_{\rm e}(E,Z)$, injected into \hho\ and $\ce{O2}$ ice calculated with SRIM, we find 
\begin{equation}
    \frac{\rm dose}{\rm eV/atom} = \left(\frac{t}{\rm yr}\right) \left(\frac{\zeta}{10^{-17}\,\rm s^{-1}}\right)\times \begin{cases}
        1.82\times10^{-8}\ \text{for \ce{H2O} ice},  \\ 
        4.14\times 10^{-8}\ \text{for \ce{O2} ice}, 
    \end{cases} 
\end{equation}
where the dose is given per atom in the target icy molecular species, \hho\ and $\ce{O2}$. 
In the following sections, the simulated results are demonstrated as functions of ion fluence. 

\section{Simulation results}
\label{sec:res}

\subsection{Experimental data of proton-irradiated pure \hho\ and \oo\ ice}
\label{sec:exp_data}

\subsubsection{Pure \hho\ ice} \label{sec:exp_data_H2O}
Using the chemical model and reaction network introduced in Section \ref{sec:model&network}, we can simulate the chemistry of pure thick \hho\ ice irradiated by 200 keV protons at 16 K and 77 K. 
As in SVHC19, our model results are compared with the experiments conducted by \citet{Gomis_Hydrogen_2004}, who obtained a maximum \hoohbyhho\ abundance ratio of 1.2\% at 16 K and 0.4\% at 77 K. 
If the column density of \hho\ is assumed to be constant throughout the experiment, the experimental \hoohbyhho\ as a function of ion fluence can be fitted by: 
\begin{equation}
    \frac{\rm [H_2O_2]}{\rm [H_2O]} = \left( \frac{\rm [H_2O_2]}{\rm [H_2O]} \right)_{\rm max} \left( 1-{\rm e}^{-\sigma_{\rm d}F} \right), 
\end{equation}
where $F$ is the fluence and $\sigma_{\rm d}$ is the destruction cross-section of \hooh. 
The results of \citet{Gomis_Hydrogen_2004} can be fitted with $\sigma_{\rm d}=9\times10^{-15}\rm \ cm^2$ and $1.6\times10^{-14}\rm \ cm^2$ at 16 K and 77 K, respectively. 
We also estimated from the column density evolution of water ice upon irradiation presented in Fig. 1 of \citet{Gomis_Hydrogen_2004} that $\lesssim 8\%$ of the initial \hho\ is destroyed by 200 keV $\rm He^+$ ion at 16 K at large fluence, providing an upper limit to the total number of species which can be produced.
The expected destruction of \hho\ in a proton-irradiated \hho\ experiment can be obtained by scaling this result from the $\rm He^+$ experiment by the ratio of the stopping powers of proton ($1.93\times 10^{-14}\rm \ eV\ cm^2\ molecule^{-1}$) and $\rm He^+$ ($4.98\times 10^{-14}\rm \ eV\ cm^2\ molecule^{-1}$) in \hho\ ice. 
We thus evaluate that $\lesssim 3.1\%$ of the initial \hho\ should be destroyed after irradiation of 200 keV protons at same fluences at 16 K. 

\par

A similar experiment, with pure \hho\ ice exposed to 100 keV protons at 20 K and 80 K, was conducted by \citet{Loeffler_Synthesis_2006} who concluded that the maximum \hoohbyhho\ is 0.75\% at 20 K and 0.14\% at 80 K. 
Although the temperature difference may contribute to the resulting maximum \hoohbyhho, we can roughly assume that the typical uncertainty in experimental results is a factor of 3. 

\par

\hoohbyhho\ is the only value shown in \citet{Gomis_Hydrogen_2004} and \citet{Loeffler_Synthesis_2006}. 
However, experiments with pure \hho\ ice exposed to other ionizing radiation could also hint the product of proton-irradiated \hho\ ice which we focus on. 
Exposed to UV photons composed of a Lyman-$\alpha$ (121.6 nm) and molecular \hh\ emission bands (130--165 nm), \hho\ ice can produce \oo\ and \hooh\ with $[\ce{O2}]/[\ce{H2O2}]\sim 0.7$ at 20 K \citep{Bulak_Quantification_2022}. 
When \hho\ ice is irradiated by 5 keV electrons, \hh, \oo\ and \hooh\ were found to be formed with $[\ce{H2}]:[\ce{H2O2}]:[\ce{O2}]\approx 160:13:1$ at 12 K \citep{Zheng_Temperature_2006}. 
Although the exact amounts of the products are different in \hho\ ice irradiated by protons, these experiments show that a significant amount of \hh\ and \oo\ are expected to form together with \hooh. 

\subsubsection{Pure \oo\ ice} \label{sec:exp_data_O2}
Pure \oo\ ice is the simplest system for proton irradiation experiment and simulation, with only three species O, \oo, and \ooo\ to consider. 
An experiment of \oo\ ice exposed to 5 keV protons was conducted by \citet{Ennis_formation_2011} at 12 K. 
They obtained a steady state \ooo\ column density of $1.5\times 10^{16}\rm \ cm^{-2}$, with an initial \oo\ column density of $3.8\times 10^{18}\rm \ cm^{-3}$. 
In this section, we will compare our simulation results with this experimental value $[\ce{O3}]/[\ce{O2}]_0=0.395\%$ ($[\ce{O2}]_0$ refers to the initial abundance of \oo\ ice) with an uncertainty of a factor of 3. 
Although \citet{Ennis_formation_2011} did not fit the fluence-dependent evolution of \ooo\ ice, they reported that the \ooo\ column density reached the steady state after a irradiation time of $2.5\times 10^4$ s, which corresponds to a fluence of $\sim 7.6\times 10^{15}\rm \ ions\,cm^{-2}$. 

\par

We note that \citet{Baragiola_Solid-state_1999} conducted a similar experiment, irradiating pure \oo\ ice with 100 keV protons at 5 K. 
The density of \ooo\ was reproduced by simulations (SVHC19). 
However, \citet{Baragiola_Solid-state_1999} did not report the density of \oo\ in their experiment. 
In SVHC19, they adopted the initial \oo\ density from \citet{Horl_Structure_1982}. 
Since this value is not directly obtained from the experiment of \citet{Baragiola_Solid-state_1999}, it may bring extra uncertainty to the analysis. 
Therefore, we choose the experiment of \citet{Ennis_formation_2011} as reference to be more consistent.

\subsection{Results of our best-fit model}
\label{sec:res_best}

\subsubsection{Pure \hho\ ice} \label{sec:res_best_H2O}

\begin{figure}
\centering
\includegraphics[width=0.99\hsize]{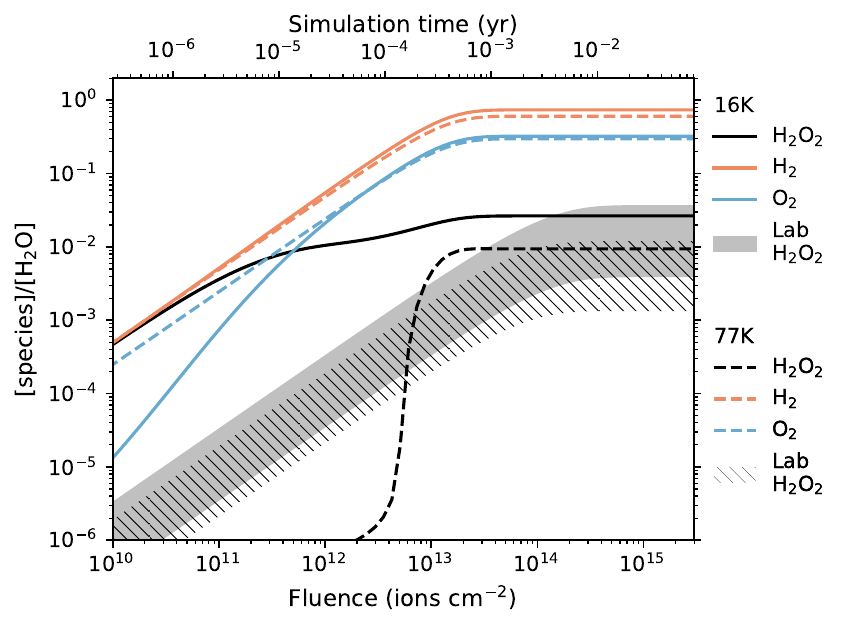}
\caption{Simulation results of abundance ratios \hoohbyhho\ (black lines), \hhbyhho\ (orange lines), and \oobyhho\ (blue lines) with our model (model A) at 16 K (solid lines) and 77 K (dashed lines). 
The gray shaded and line filled regions show the experimental results of \hoohbyhho\ by \citet{Gomis_Hydrogen_2004} at 16 and 77 K, respectively, with an assumed uncertainty of a factor of 3. 
}
\label{fig:res_best}
\end{figure}

The simulation results of proton-irradiated pure \hho\ ice at 16 K and 77 K with our model (referred to as model A hereafter), as describes in Section \ref{sec:model&network}, are shown in Figure \ref{fig:res_best}. 
The abundance ratios \hoohbyhho, \hhbyhho\ and \oobyhho\ rise monotonically with the ion fluence when the fluence is $\lesssim 3\times 10^{13}\rm \ ions\ cm^{-2}$, and reach the steady states afterwards. 
The steady state \hoohbyhho\ are 2.65\% at 16 K and 0.941\% at 77 K, roughly consistent with the experimental results. 
Our model also produces significant amounts of \hh\ and \oo, contrary to SVHC19, with $[\ce{H2}]/[\ce{H2O}]=74.3\%$ and $[\ce{O2}]/[\ce{H2O}]=32.5\%$ at 16 K, and $[\ce{H2}]/[\ce{H2O}]=60.8\%$ and $[\ce{O2}]/[\ce{H2O}]=29.9\%$ at 77 K. 
However, we note that the fluence required to reach the steady state is smaller than the experimental results by approximately one order of magnitude. 
In addition, the model destroys more \hho\ (43\% at 16 K and 41\% at 77 K) than the experiment, resulting in high steady-state \hhbyhho\ and \oobyhho\ ratios. 

\par

In order to understand the chemical processes, we looked at the reactions involved. 
At 16 K, the major formation reactions of \hooh\ are the barrierless non-diffusive reactions: 
\begin{align}
    & \ce{H + O2H -> H2O2}, \tag{R7} \label{reac:H+O2H89} \\
    & \ce{O2H + O2H -> H2O2 + O2}, \tag{R8} \label{reac:O2H+O2H89} \\
    & \ce{O^* + H2O -> H2O2}, \tag{R9} \label{reac:O^*+H2O89} \\
    & \ce{OH^* + H2O -> H2O2 + H}. \tag{R10} \label{reac:OH^*+H2O89}
\end{align}
At 77 K, \hooh\ is mainly formed via reaction \ref{reac:O2H+O2H89}, with minor contributions from reactions \ref{reac:O^*+H2O89} and 
\begin{equation}
    \ce{O3^* + H2O -> H2O2 + O2} \tag{R11} \label{reac:O3*+H2O89}. 
\end{equation}
At both temperatures, \hooh\ is destroyed by the non-diffusive reaction: 
\begin{equation}
    \ce{OH + H2O2 -> H2O + O2H} \tag{R12}.  \label{reac:OH+H2O289}. 
\end{equation}
Note that this destruction pathway was also identified as the major one at 77 K by SVHC19, but we cannot compare for the formation or destruction at lower temperature as the main reactions were not given by the authors. 

\subsubsection{Pure \oo\ ice} \label{sec:res_best_O2}

\begin{figure}
\centering
\includegraphics[width=0.99\hsize]{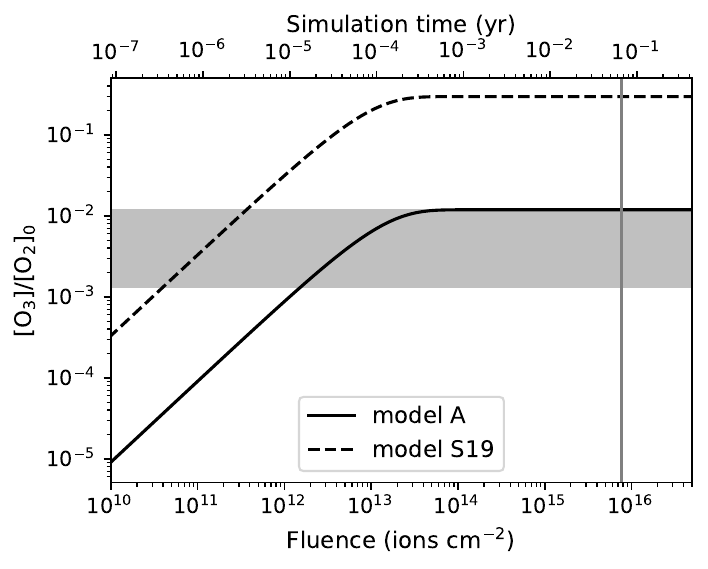}
\caption{Simulation results of \ooobyoo\ irradiated by 5 keV protons at 12 K based on model A (solid line) and model S19 (dashed line). 
The gray shaded region shows the range of steady state \ooobyoo\ obtained by \citet{Ennis_formation_2011} with an uncertainty of a factor of 3. 
The vertical gray line shows the fluence of $7.6\times 10^{15}\rm \ ions\, cm^{-2}$ which is needed to reach steady state in the experiment of \citet{Ennis_formation_2011}. 
}
\label{fig:res_O2}
\end{figure}

The simulation results of proton-irradiated pure \oo\ ice at 12 K with model A are shown in Figure \ref{fig:res_O2}. 
The steady-state \ooobyoo\ is 1.19\%, roughly consistent with the experimental results. 
However, the fluence required to reach the steady state is $\sim 2$ orders of magnitude smaller than the experimental values. 
In this simulation, \ooo\ is mainly formed via non-diffusive reaction: 
\begin{equation}
    \ce{O^* + O2 -> O3} \tag{R13} \label{reac:O*+O289}
\end{equation}
and the quenching of $\ce{O3^*}$. 
It is mainly destroyed by radiolysis, with minor contributions from reactions: 
\begin{align}
    & \ce{O2^* + O3 -> 2O2 + O}, \tag{R14} \label{reac:O2^*+O389} \\
    & \ce{O^* + O3 -> 2O2}. \tag{R15} \label{reac:O*+O389} 
\end{align}

\subsection{Model results using SVHC19 network}
\label{sec:res_S19}

As shown in Section \ref{sec:network}, the chemical network of model A is slightly different from that used in SVHC19. 
Hereafter we name the model based on their network as S19. 
We briefly summarize the differences below: 
\begin{itemize}
    \item[1.] Branching ratios of the $\ce{H + O2H}$ reactions: we set them to 0.2, 0.2, and 0.6 for the pathways that produce $\ce{OH + OH}$, $\ce{O2 + H2}$, and \hooh, respectively, according to \citet{Cuppen_Water_2010,Lamberts_Water_2013}, while in model S19, these values are 0.0194, 0.0857, and 0.894, respectively.  
    \item[2.] Activation barriers and branching ratios of the $\ce{H + H2O2}$ reactions: we adopt 0.5 and 3400 K for the branching ratio and activation barrier for reaction $\ce{H + H2O2 -> H2O + OH}$, respectively, and 0.5 and 4200 K for reaction $\ce{H + H2O2 -> O2H + H2}$ \citep{Ellingson_Reactions_2007,Lamberts_Water_2013,Lamberts_Quantum_2016,Lamberts_Influence_2017,Lu_Rate_2018,Lu_Theoretical_2019}, while in S19, the values are 0.999 and 1400 K for the former reaction and 0.0006 and 1900 K for the latter reaction. 
    \item[3.] Activation barriers of the $\ce{O + O2 -> O3 / O3^*}$ reactions: we set all of them to 300 K, while S19 adopted 180 K for $\ce{O + O2 -> O3}$ and 0 K for $\ce{O + O2 -> O3^*}$. We modify this in order to reduce the formation of \ooo\ and $\ce{O3^*}$, and this value is within the range estimated from the experiment of \citet{Benderskii_Diffusion-limited_1996}. 
    \item[4.] Branching ratios of the $\ce{O^* + O2}$ reactions: we set them to be 0.05, 0.90 and 0.05 for the reactions that produce $\ce{O3}$, $\ce{O + O2}$, and $\ce{O3^*}$, respectively, while in S19, the values are all 0.333. This modification is to reduce the formation of \ooo\ and $\ce{O3^*}$. 
    \item[5.] Activation barrier of the $\ce{OH + H2O2 -> H2O + O2H}$ reaction: we set it to 220 K, while S19 used 760 K. We modify this barrier to increase the destruction of \hooh, and our value is reasonable according to \citet{Atkinson_Evaluated_2004,Buszek_Effects_2012}.
    \item[6.] Reaction pathways of the $\ce{O2H + O2H}$ reaction: we add a pathway to produce $\ce{H2O + O3}$ \citep{Zhang_Water-catalyzed_2011} with a branching ratio of 0.1 in addition to the $\ce{H2O2 + O2}$ pathway in S19. \citet{Silva_Molecular_2025} found that the branching ratio of the $\ce{H2O + O3}$ pathway is 0.19. Here we adjust it to 0.1 to better fit the experiments. Adding this pathway reduces the formation of \hooh, especially at 77 K. 
\end{itemize}
There are many uncertainties in branching ratios, reaction activation energies, and even chemical pathways. For activation energies, we checked all the references in the NIST chemical kinetics database\footnote{\url{https://kinetics.nist.gov/kinetics/}}, showing the variation in the values from different experiments or theoretical calculations. Note that the exact values here in model A were tuned to obtain a better agreement with the experiences (see also section~\ref{sec:compChris}). Here we use model S19 to conduct the simulation of proton-irradiated \hho\ and \oo\ ice to justify how these changes in activation barriers and branching ratios affect the simulation results. 

\subsubsection{Pure \hho\ ice} \label{sec:res_S19_H2O}
\begin{figure}
\centering
\includegraphics[width=0.99\hsize]{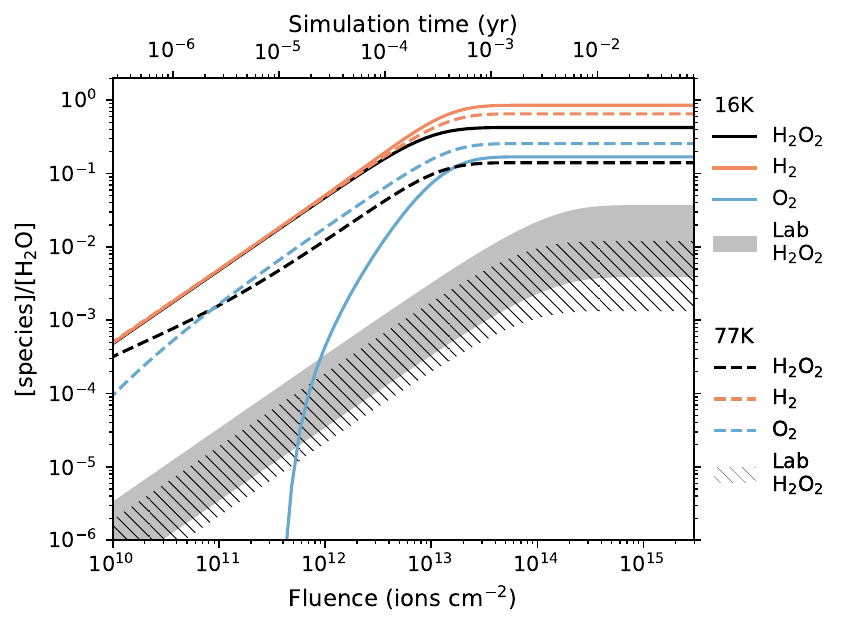}
\caption{Same as Figure \ref{fig:res_best} but for model S19. 
}
\label{fig:res_S19}
\end{figure}

The simulation results of proton-irradiated pure \hho\ ice at 16 K  and 77 K with model S19 are shown in Figure \ref{fig:res_S19}. 
Similar to model A, the steady state is reached at $\sim 3\times 10^{13}\rm \ ions\ cm^{-2}$, earlier than the experiment. 
However, model S19 results in $[\ce{H2O2}]/[\ce{H2O}]=42.7\%$ and 16 K and 14.1\% at 77 K, both significantly higher than the experimental results. 
It is likely that the higher \hoohbyhho\ abundance ratio in model S19 than model A is due to the higher activation barrier of reaction \ref{reac:OH+H2O289}, which is 760 K in model S19 and 220 K in model A. 
At 16 K, \hooh\ is mainly destroyed by the radiolysis reactions instead of \ref{reac:OH+H2O289} which is the major destruction reaction in model A. 
At 77 K, although the higher dust temperature allows for overcoming the activation barrier, the reaction rate of \ref{reac:OH+H2O289} is still lower in model S19 than that in model A, and the radiolysis reactions of \hooh\ contribute more to the its destruction in model S19. 
The major formation reactions of \hooh\ at 16 K are non-diffusive reactions: 
\begin{equation}
    \ce{OH + OH -> H2O2}, \tag{R16} \label{reac:OH+OH89}
\end{equation}
\ref{reac:H+O2H89}, \ref{reac:O3*+H2O89}, \ref{reac:OH^*+H2O89}, and \ref{reac:O^*+H2O89}. 
At 77 K, \hooh\ is mainly formed via reaction \ref{reac:O2H+O2H89} with minor contributions from reactions \ref{reac:O3*+H2O89} and \ref{reac:O^*+H2O89}. 

\subsubsection{Pure \oo\ ice} \label{sec:res_S19_O2}
The simulation results of proton-irradiated \oo\ ice is shown in Figure \ref{fig:res_O2}. 
Generally, the results of model S19 follows the similar shape as those of model A, but the steady state \ooobyoo\ is 29.6\%, much higher than the experimental results. 
The activation barriers of reactions: 
\begin{align} 
    & \ce{O + O2 -> O3} \tag{R17} \label{reac:O+O2->O389} \\ 
    & \ce{O + O2 -> O3^*} \tag{R18} \label{reac:O+O2->O3*89}
\end{align}
are lower in model S19 than in model A, which boost the formation of \ooo.  
Another difference between models A and S19 is the branching ratios of reactions: 
\begin{equation}
\ce{O^* + O2 ->} \begin{cases}
\ce{O3} \\
\ce{O + O2} \\
\ce{O3^*}
\end{cases}
\tag{R19}
\end{equation}
which are all 0.333 in model S19, and 0.05, 0.90 and 0.05 in model A. 
This further increases the steady-state \ooobyoo\ in model S19. 
The destruction reactions of \ooo\ in model S19 are: 
\begin{equation}
    \ce{O3^* + O3 -> 3O2}, \tag{R20} \label{reac:O3^*+O389}
\end{equation}
radiolysis reactions, \ref{reac:O*+O389} and \ref{reac:O2^*+O389}. 

\subsection{Sensitivity to the CR ionization rates}
\label{sec:res_zeta}

\begin{figure}
\centering
\includegraphics[width=0.99\hsize]{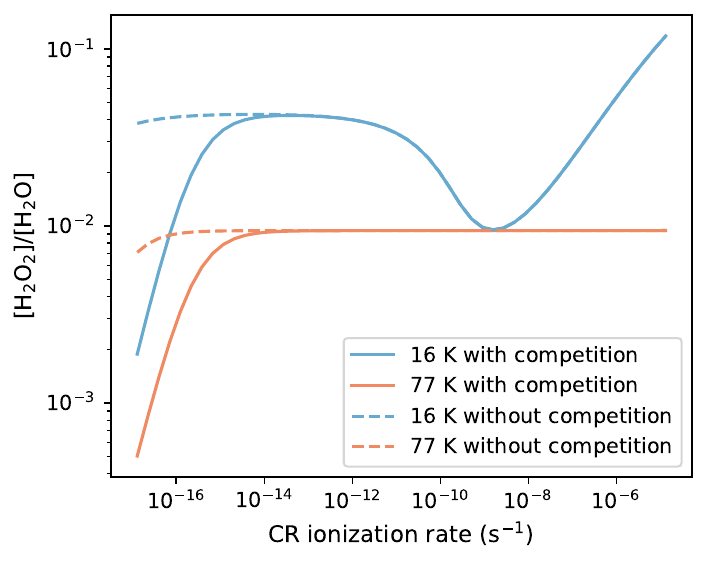}
\caption{Simulation results of the steady-state \hoohbyhho\ abundance ratio in model A as a function of the CR ionization rate. 
The results at 16 K and 77 K are shown in blue and orange lines, respectively. 
The solid and dashed lines show the results with and without the reaction-diffusion competition, respectively. 
}
\label{fig:res_zeta-ratio_H2O}
\end{figure}

\begin{figure}
\centering
\includegraphics[width=0.99\hsize]{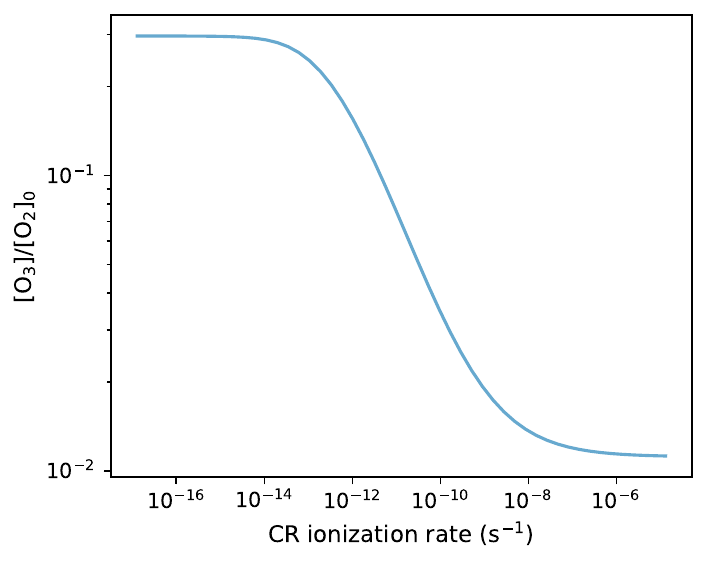}
\caption{Simulation results of the steady-state \ooobyoo\ abundance ratio in model A as a function of the CR ionization rate. 
}
\label{fig:res_zeta-ratio_O2}
\end{figure}

Most experiments study the \hho\ radiolysis effect with a CR flux of $\sim 10^9$--$10^{12}\rm \ ions\ s^{-1}\ cm^{-2}$ \citep[][]{Moore_IR_2000,Gomis_Hydrogen_2004a,Gomis_Hydrogen_2004,Mejia_Swift_2022}, corresponding to the CR ionization rates of $\sim 10^{-9}$--$10^{-6}\rm \ s^{-1}$, due to the limitation of the instruments and the experiment time. 
Their results show that the steady-state \hoohbyhho\ abundance ratio is, though not perfectly, independent on the CR flux in the experiment \citep[][]{Moore_IR_2000,Gomis_Hydrogen_2004a,Gomis_Hydrogen_2004,Mejia_Swift_2022}. 
In Figure \ref{fig:res_zeta-ratio_H2O} we show the simulation results of the steady-state \hoohbyhho\ abundance ratio in model A as a function of the CR ionization rate at 16 K and 77 K. 
At both temperatures, \hoohbyhho\ increases with the CR ionization rates when $\zeta \lesssim 10^{-14}\rm \ s^{-1}$. 
At higher $\zeta$, the model at 77 K results in a constant \hoohbyhho, while the model at 16 K shows a complicated variation in \hoohbyhho. 
It decreases to a local minimum at $\zeta\sim 10^{-9}$--$10^{-8}\rm \ s^{-1}$, and then increases monotonically at higher $\zeta$. 
Though, the \hoohbyhho\ ratios in the range $10^{-9}{\rm\ s^{-1}} < \zeta < 10^{-6}{\rm\ s^{-1}} $ vary roughly within a factor of 2--3 around $\sim 2\%$, lower than the experimental uncertainties. 

\par

Similarly, we show the \ooobyoo-$\zeta$ plot of model A in Figure \ref{fig:res_zeta-ratio_O2}. 
The \ooobyoo\ abundance ratio decrease with the CR ionization rate. 
But within the experimental range of CR flux, i.e. $10^{-9}{\rm\ s^{-1}} < \zeta < 10^{-6}{\rm\ s^{-1}} $, the variation of the ratio is rather small. 
Therefore, we conclude that model A successfully reproduce the constant steady-state abundance ratios within the experimental range of CR flux in both the \hho\ and \oo\ experiments.

\subsection{Sensitivity to $G$-values of \hho\ radiolysis}
\label{sec:res_lowG}

\begin{figure}
\centering
\includegraphics[width=0.99\hsize]{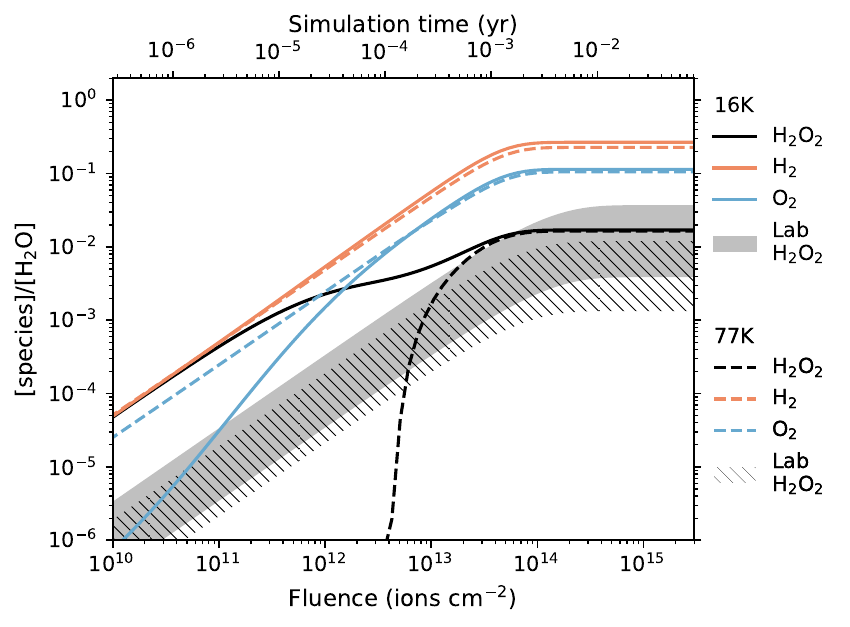}
\caption{Same as Figure \ref{fig:res_best} but for model A with the $G$-values of \hho\ radiolysis processes lowered by an order of magnitude. 
}
\label{fig:res_best_lowG}
\end{figure}

The exact values of the $G$-values listed in Table \ref{tab:radiolysis_reactions} actually depend on several factors and assumptions such as the intermediate pathways that are considered in the radiolysis process. 
In \citet{Shingledecker_Cosmic-Ray-driven_2018}, they adopted 3.704 for type R1 radiolysis, and 1.747 for types R2 and R3. 
On the other hand, SVHC19 adopted 0.257 for type R1 and 0.158 for types R2 and R3. 
The higher values are the number of the dissociated \hho\ after every 100 eV deposited, while the lower values are adopted to provide ``effective'' $G$-values that include implicitly reformation processes of \hho\ to compensate for those potentially underconsidered in the model. 
Since our model ignores the chemical effects of charged species, we may have missed some pathways in the radiolysis process, which causes an overestimate on the $G$-values. 

\par

Our model A overestimates the destruction of \hho\ (see Section \ref{sec:res_best_H2O}). 
In this section, we test the sensitivity of the simulation results to the uncertainties in $G$-values of \hho\ radiolysis by lowering them by an order of magnitude to see whether this can significantly reduce the destruction of \hho while maintaining a significant amount of \hooh\ production. 
The results are shown in Figure \ref{fig:res_best_lowG}. 
At 16 K, the steady-state \hoohbyhho\ is 1.71\%, still consistent with the experimental values. 
In this case, 22\% of the initial \hho\ is destroyed, which is lower than that in model A and closer to the experimental values, resulting in steady-state \hhbyhho\ and \oobyhho\ of 26.8\% and 11.4\%, respectively. 
At 77 K, the steady-state \hoohbyhho\ is 1.65\%, slightly higher than that in model A and the experimental value. 
About 20\% of the initial \hho\ is destroyed. 
Our results show that by lowering the $G$-values of the \hho\ radiolysis processes, we can reduce the destruction of \hho\ and make the simulation results closer to the experimental values, while the \hoohbyhho\ is still roughly consistent with the experiments.

\subsection{Effects of reaction-diffusion competition}
\label{sec:res_compe}
As mentioned in Section \ref{sec:model}, the reaction-diffusion competition mechanism was introduced into the nautilus code to make it more likely for the reaction with finite activation barriers to take place \citep{Chang_Gas-grain_2007,Garrod_Formation_2011,Ruaud_Gas_2016}. 
On the other hand, the model of SVHC19 did not include this competition. 
In Figure \ref{fig:res_zeta-ratio_H2O}, we also show the \hoohbyhho-$\zeta$ plot in model A without the reaction-diffusion competition to compare with the case with the competition. 
At both 16 K and 77 K, the models without the competition result in higher \hoohbyhho\ at $\zeta \lesssim 10^{-14}\rm \ s^{-1}$ are compared with the case with the competition. 
At $\zeta \gtrsim 10^{-14}\rm \ s^{-1}$, the results of the two cases are consistent. 

\par

To explore the different \hoohbyhho\ ratio at $\zeta \lesssim 10^{-14}\rm \ s^{-1}$ between the models with and without the competition, we look at the major formation and destruction reactions of \hooh\ at low $\zeta$, and find that they are the same as the reactions at high $\zeta$, i.e. formation via \ref{reac:H+O2H89}, \ref{reac:O2H+O2H89}, \ref{reac:O^*+H2O89}, \ref{reac:OH^*+H2O89} and \ref{reac:O3*+H2O89}, while destruction by \ref{reac:OH+H2O289}. 
Note that all the formation reactions are without activation barriers, while the destruction reaction, \ref{reac:OH+H2O289}, has a barrier of 220 K in model A. 
Therefore, inclusion of the reaction-diffusion competition can increase the destruction rate of \hooh\ and thus lead to lower \hoohbyhho. 
On the other hand, at higher $\zeta$, the abundance of OH increases significantly due to the faster radiolysis of \hho\ that boost the formation of OH. 
The large amount of OH destroys \hooh\ quickly, so the steady-state \hoohbyhho\ is determined by the formation rate of \hooh\ instead of its destruction. 
Therefore, the reaction-diffusion competition does not affect the \hoohbyhho\ any more at high $\zeta$. 
This explanation may also be supported by the fact that at low $\zeta$, the destruction of OH is only via reaction \ref{reac:OH+H2O289}, while at high $\zeta$, reactions with H and \ooh\ also contribute to the destruction of OH in addition to reaction \ref{reac:OH+H2O289}. 
This means that OH is lacking at low $\zeta$ and is consumed totally by \hooh, but at high $\zeta$, OH is so abundant that it can not only consume the \hooh\ molecules, but also take part in the destruction of H and \ooh. 
Such abundant OH at high $\zeta$ makes the reaction \ref{reac:OH+H2O289} no longer limited by the activation barrier, so the reaction-diffusion barrier does no longer affect the destruction of \hooh. 

\par

We note that the inclusion of the reaction-diffusion competition does not influence the simulation results in the \oo\ experiment. 
So we do not show these results in Figure \ref{fig:res_zeta-ratio_O2}.

\subsection{Effects of non-diffusive chemistry}
\label{sec:res_non-diff}

\begin{figure}
\centering
\includegraphics[width=0.99\hsize]{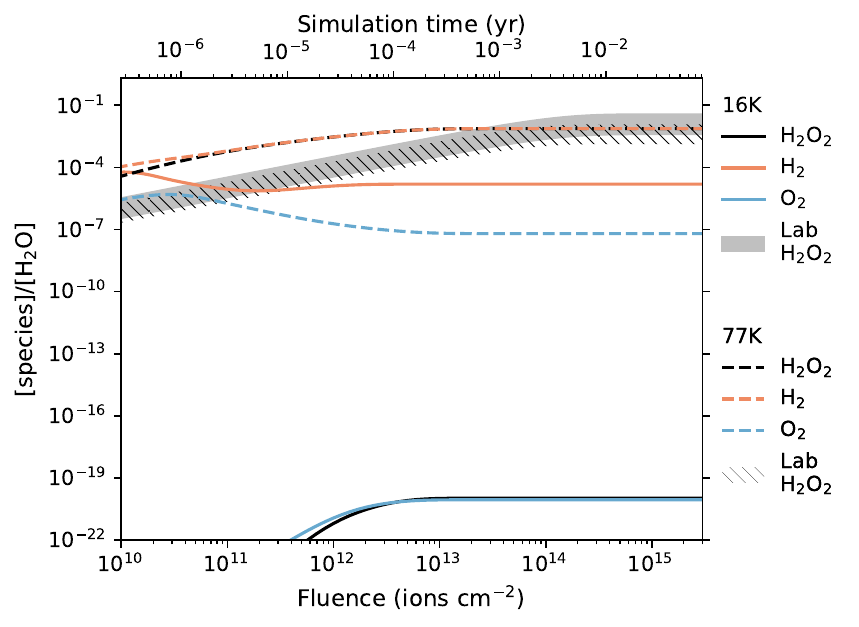}
\caption{Same as Figure \ref{fig:res_best} but for model A with all non-diffusive reactions removed. 
}
\label{fig:res_best_diff}
\end{figure}

We notice that all the reactions that lead to the formation and destruction of \hooh\ are non-diffusive reactions. 
In Figure \ref{fig:res_best_diff}, we show the simulation results of model A with all the non-diffusive reaction removed. 
The steady-state \hoohbyhho\ abundance ratio is extremely low ($\sim 10^{-20}$) at 16 K but 0.736\% at 77 K. 
This means that at 16 K, non-diffusive chemistry is crucial to the formation of \hooh. 
However, at 77 K, the \hoohbyhho\ ratio does not differ significantly from the result of model A (0.941\%). 
In this case, \hooh\ forms from the reaction 
\begin{equation}
    \ce{H + O2H -> H2O2}. \tag{R21} \label{reac:H+O2H14}
\end{equation}
The formation of \ooh\ relies on the reaction between two O-bearing species, and is thus slow at 16 K but fast at 77 K because higher temperature is needed to allow the O-bearing species diffuse on the dust grain before they react with each other.

\subsection{Sensitivity to the desorption energies of suprathermal species}
\label{sec:res_ED}

\begin{figure}
\centering
\includegraphics[width=0.99\hsize]{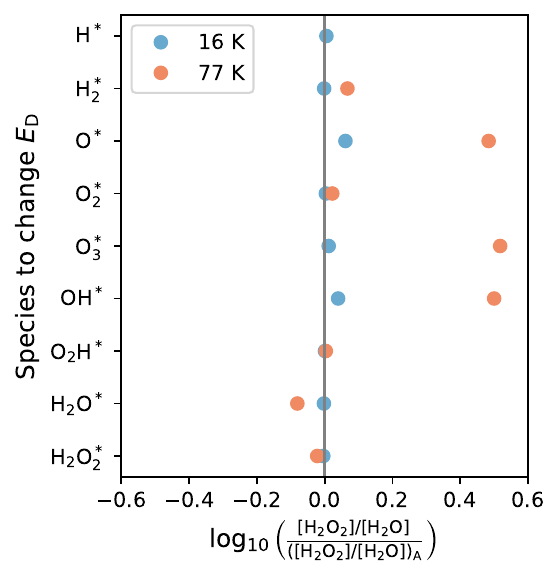}
\caption{Logarithm of the ratio between the simulated \hoohbyhho\ after lowering the desorption energy of a specific suprathermal species by an order of magnitude and the simulated \hoohbyhho\ in model A. 
The results at 16 K are shown in blue and those at 77 K are shown in orange. 
The vertical grey line shows the positions where \hoohbyhho\ is not changed after changing the desorption energy. 
Note that the ratio when changing the $E_{\rm D}$ of $\rm H^*$ at 77 K is $-3.86$, far outside the range of the x axis, and is thus not plotted in the figure. 
}
\label{fig:res_ED}
\end{figure}

The desorption energies ($E_{\rm D}$) of mantle species may also affect the result of the simulations. 
For diffusive reactions, the rate coefficient depends on $\exp{(-E_{\rm diff}/T_{\rm ice})}$ where $E_{\rm diff}=0.7E_{\rm D}$ except for H and \hh. 
It also depends on the characteristic vibrational frequencies of the reactants with $\nu \propto \sqrt{E_{\rm D}}$. 
For non-diffusive reactions, the rate coefficient depends only on $\nu$. 
The $E_{\rm D}$ of the suprathermal species are unknown due to the lack of experiments and are assumed to be the same as those of the corresponding thermal species. 
In this section, we explore the possible effect of changing the $E_{\rm D}$ of the suprathermal species. 
We conduct several tests, arbitrarily lowering the $E_{\rm D}$ of one suprathermal species at a time by an order of magnitude based on model A, since the suprathermal species are more active than the thermal species. 
We record the variation in \hoohbyhho\ compared with model A, and the results are shown in Figure \ref{fig:res_ED}. 
The results show that the \hoohbyhho\ ratio is not sensitive to the change in the $E_{\rm D}$ of the suprathermal species at 16 K. 
On the other hand, at 77 K, the \hoohbyhho\ is enhanced by a factor of $\sim 3$ after lowering the $E_{\rm D}$ of $\rm O^*$, $\rm OH^*$ and $\rm H_2O^*$. 
In addition, changing the $E_{\rm D}$ of $\rm H^*$ causes a decrease of \hoohbyhho\ to $\sim 10^{-4}$ times the value in model A. 
Therefore, the simulation result is sensitive to the values of $E_{\rm D}$ of some suprathermal species under specific conditions, but relevant experiment is still scarce.

\section{Discussion} \label{sec:disc}

\subsection{Comparison with \citet{Shingledecker_Simulating_2019}} \label{sec:compChris}
Our model A successfully reproduces the steady-state \hoohbyhho\ in the \hho\ experiment and the \ooobyoo\ in the \oo\ experiment, although the model underestimate the ion fluence required to reach the steady state. 
However, when we switch back to the network of SVHC19, we do not reproduce the experiments contrary to them while using the same chemical network and formalism of radiolysis, quenching of suprathermal species, and non-diffusive chemistry. 
The S19 model always over-produce \hooh\ in the \hho\ experiment (see Figure \ref{fig:res_S19}) and \ooo\ in the \oo\ experiment (see Figure \ref{fig:res_O2}). 
As stated in section~\ref{sec:network}, our models include a number of differences. The main difference is probably that SVHC19 included the modified rate equations capable of taking into account stochastic effects in surface chemistry as proposed by \citet{Garrod_new_2008}. However, these modified rates should only influence the model results only in the limit accretion case (when the number of reactants on the surface is close to 1), which is not our case here. 
We also did not use the same electronic stopping power as SVHC19, which resulted in different radiolysis rates. In addition, there may be different treatments or slightly different surface parameters between the two models, which are difficult to assess. Note however that with their best model, SVHC19 produced negligible amounts of \oo\ when simulating \hho\ radiolysis. 
This may be because they used the ``effective'' G-values, which is lower than the destruction G-values, that take into account the reformation processes of \hho\ that are not considered in the model.

\par

In order to make the simulated results closer to the experimental ones, several modifications are made on the reaction network based on the one used in SVHC19, as introduced in Section \ref{sec:res_S19}. 
By increasing the activation barriers of the $\ce{O + O2 -> O3 / O3^*}$ reactions and the branching ratios of the $\ce{O^* + O2}$ reactions, we produce less \ooo\ (and $\ce{O3^*}$) in the \oo\ experiment compared with the network of SVHC19. 
By lowering the activation barrier of reaction \ref{reac:OH+H2O289} from 760 K in model S19 to 220 K in model A, we manage to make it the major destruction reaction of \hooh, and at the same time, we add a new pathway of the $\ce{O2H + O2H}$ reaction to reduce the formation rate of \hooh. 
All these changes reduce the \hoohbyhho\ and \ooobyoo\ ratios from model S19 and make model A consistent with the experiments. 

\par

The comparison between models A and S19 strongly stresses the importance of the activation barriers and the branching ratios of the reactions in the chemical network. 
In such a small network to reproduce the radiolysis experiment of pure ices, the exact values of the activation energies and the branching ratios of some reactions can completely change the steady-state abundances. 
For example, by adding the reaction pathway of $\ce{O2H + O2H -> H2O + O3}$ with a branching ratio of 0.1 and reducing the branching ratio of reaction $\ce{O2H + O2H -> H2O2 + O2}$ (\ref{reac:O2H+O2H89}) to 0.9, the steady-state \hoohbyhho\ is reduced by a factor of $\sim 8$ (from 7.11\% to 0.941\%, not shown in Section \ref{sec:res}). 
Many of the activation barriers and branching ratios, if not all, are poorly or completely unknown for surface reactions. Indeed, the values for which we have estimates and references have been obtained for the gas phase at room temperature, and can be much lower in the ices. As an example, the activation energy for CO + O in the gas phase at room temperature is between 2500 and 3000 K \citep{Talbi_interstellar_2006} while it is only 627 K in the ices \citep{Minissale_CO2_2013}.
As such, we did not test them all to get a prefect agreement with the experiments. 
In addition, the desorption energies of the suprathermal species, which are also poorly constrained by experiments, may also affect the simulation results.

\subsection{Uncertainties in $G$-values of \hho\ radiolysis}
In this study, we only consider three processes of radiolysis (see Section \ref{sec:model}) and ignore the production of charged species on dust grains. 
It is generally believed that grains in MCs are negatively charged, and molecular ions adsorbed or formed on dust grains will soon recombine with the electrons on the dust surface \citep{Umebayashi_Recombination_1980,Aikawa_Grain_1999}.
However, it is not understood what the case is in the bulk of the ice, namely, whether molecular ions can react with other species in the bulk. 
Kinetic simulation of ion chemistry on the surface and in the bulk is challenged by the lack of theoretical calculations and experiments of negatively charged dust grains \citep{Cui_Exploring_2024}. 
We note that charged particles, such as $\ce{H2O+}$, $\ce{H3O+}$, $\ce{OH-}$ and $\ce{e-}$ \citep{Muroya_re-evaluation_2005,Baragiola_Radiation_2008,Teolis_Water_2017}, can be produced in the radiolysis processes and play an important role in the chemical evolution, especially for \hho\ radiolysis. 
Experiments have found that \hho\ is resilient to ion irradiation with fast reformation after dissociation \citep{LeCaer_Water_2011}. 
Our neglect of these species and processes may lead to the over-destruction, including the amount and rate of the destruction, of \hho\ in model A. 
By arbitrarily lowering the $G$-values of \hho\ radiolysis by an order of magnitude (see Section \ref{sec:res_lowG}), i.e. taking into consideration an "effective" destruction rate lower than the primary destruction induced by the impinging ion, we can make the simulation results closer to the experiments especially regarding the fluence required to reach the steady state. 
Therefore, the lack of details about the intermediate pathways in the radiolysis processes may be an important reason why we could not obtain the same evolution trend of \hoohbyhho\ as the experiments.  
This test model stresses the importance of understanding these pathways and implementing them into the chemical model. 
We hence call for more studies on the ion chemistry in dust grains.

\subsection{Testing formalism for surface and bulk processes}

The chemistry in the gas-phase and at the surface of the grains are both treated using the rate-equation approach. Although such treatment is valid for gas-phase processes, it is a strong approximation for reactions at the surface of the grains and in the grain mantles. The original equations from \citet{Hasegawa_Models_1992} do not take into account the fact that species have a random walk around the grains or can stay longer at the same position, i.e. increasing the probability that a reaction takes place despite its activation energy. \citet{Chang_Gas-grain_2007} and \citet{Garrod_Formation_2011} proposed a modification of the computation of such reaction probability that is commonly called ``competition between reaction and diffusion''. 
In this study, we tested the effect of this modification in order to reproduce the experimental results for a simpler system of irradiation of pure water ices. 
The competition is initially included in model A, and we found that its removal can increase the steady-state \hoohbyhho\ at low $\zeta$ ($\lesssim 10^{-14}\rm \ s^{-1}$) because the competition facilitates the destruction of \hooh\ through reaction \ref{reac:OH+H2O289} with an activation barrier of 220 K. At higher $\zeta$, the competition has no effect on the steady-state \hoohbyhho. 

Another recent addition to astrochemical models is the treatment of non-diffusive chemistry. This process is meant to mimic the fact that when two species are produced nearby on the grains (by radiolysis for instance), they can directly react without diffusing. 
Our test models show that non-diffusive chemistry is crucial to the formation of \hooh\ in the \hho\ radiolysis experiments, especially at the lower temperature where diffusive reactions of O-bearing species are not efficient. 
SVHC19 and \citet{Jin_Formation_2020} proposed two very different formula to include the non-diffusive chemistry in models. For this work, we tried to test the difference in the model results using both of them. However, using the reduced network, the formalism by \citet{Jin_Formation_2020} could not converge. In this case, the rate of non-diffusive reactions is a sum of two terms, one of which is not proportional to the abundance of one of the reactants. This means that if one of the reactant is not produced enough, its rate of destruction can be higher than its production. In a larger network, there are many processes and this case may not happen. Non-diffusion chemistry is the most important process to reproduce experiments at low temperatures.

\subsection{Effect of the CR ionization rate}

According to Section \ref{sec:res_zeta}, model A results in nearly constant \hoohbyhho\ and \ooobyoo\ in the range of the CR flux in the experiments ($\sim 10^9$--$10^{12}\rm \ ions\ s^{-1}\ cm^{-2}$, corresponding to $\zeta\sim 10^{-9}$--$10^{-6}\rm \ s^{-1}$). 
However, the typical values of $\zeta$ in the interstellar MCs are $\sim 10^{-17}$--$10^{-16}\rm \ s^{-1}$, significantly lower than the values in the experiments. 
Figures \ref{fig:res_zeta-ratio_H2O} and \ref{fig:res_zeta-ratio_O2} show that the efficiency of radiolysis is different between these two ranges of $\zeta$. 
Although the reaction network is rather simplified, the steady-state \hoohbyhho\ and \ooobyoo\ are still not proportional to $\zeta$. 
We also note that changing the details of the network (e.g., the activation barriers and the branching ratios of specific reactions) also affects how the steady state \hoohbyhho\ evolves with $\zeta$. 
For example, if we reduce the barriers of the $\ce{O + O2}$ reactions, the sink of the \hoohbyhho\ at 16 K and $\zeta \sim 10^{-9}\rm \ s^{-1}$ in Figure \ref{fig:res_zeta-ratio_H2O} will be shallower (not shown in the figure). 
It is not clear when and why the \hoohbyhho-$\zeta$ curve can be changed when the activation barriers of some specific reactions are changed. 
Due to these complexities, it is difficult to fully interpret our results and use them to explain the observations of the real MCs.

\section{Conclusions} \label{sec:con}

Radiolysis has been proposed to be an important process to synthesize prebiotic molecules in the interstellar environment.
In this work, we try to reproduce the experiments of proton-irradiated pure \hho\ \citep{Gomis_Hydrogen_2004} and \oo\ \citep{Ennis_formation_2011} ice with the Nautilus code, a complex gas-grain astrochemical code in which we include radiolysis reactions, quenching of suprathermal species, reactions between suprathermal and thermal species, and non-diffusive reactions. 

\par

Our model A, based on the network of SVHC19 with a few modifications on the activation barriers and the branching ratios, can reproduce the steady-state \hoohbyhho\ in the \hho\ experiment and the \ooobyoo\ in the \oo\ experiment. 
It also results in almost constant \hoohbyhho\ and \ooobyoo\ ratios within the experimental CR flux, which is consistent with what was found in experiments. 
However, it underestimates the ion fluence required to reach the steady state in both the \hho\ and \oo\ experiments. 
It also overestimates the destruction of \hho, as well as the steady-state \hhbyhho\ and \oobyhho\ in the \hho\ experiment.

\par

We also conducted several tests based on model A to study how the $G$-values of the \hho\ radiolysis, the reaction-diffusion competition, the non-diffusive chemistry, and the desorption energies of the suprathermal species affect the simulation results of the \hho\ experiment. 
After we arbitrarily lower the $G$-values of the \hho\ radiolysis by an order of magnitude, to take into account potentially underconsidered reformation processes from ions produced by radiolysis in the network, the destruction of \hho\ is reduced and the simulation results are closer to experimental values.
Our neglect of the reformation processes may be the key reason why we could not reproduce the evolution trend of the studied abundance ratios. 
The reaction-diffusion competition is included in model A, and its removal will increase the \hoohbyhho\ at low $\zeta$ ($\lesssim 10^{-14}\rm \ s^{-1}$) because the competition facilitate the destruction of \hooh\ by OH. 
Non-diffusive reactions always dominate the formation and destruction of \hooh, and the removal of non-diffusive chemistry in model A strongly hampers the formation of \hooh\ at 16 K while exerts minor influence at 77 K. 
The simulation results are not sensitive to the desorption energies of the suprathermal species at 16 K, but is sensitive to the desorption energies of $\ce{H^*}$, $\ce{O^*}$, $\ce{O3^*}$, and $\ce{OH^*}$ at 77 K. 

\par

Our simulation results show that the steady-state \hoohbyhho\ and \ooobyoo\ in radiolysis experiments can be reproduced by fine-tuning the chemical model. 
However, we still call for more experiments to determine the intermediate processes in the radiolysis, especially the ion chemistry in the ice bulk, as well as the activation barriers and the branching ratios of the reactions in the network, which is crucial to our understanding of the radiolysis process via the chemical model. 

\par

The modified \texttt{Nautilus} code, with the inclusion of non-diffusive chemistry, radiolysis, and some other processes, will be introduced in a forthcoming paper (Wakelam et al. in prep.) and will be made available at \url{https://forge.oasu.u-bordeaux.fr/LAB/astrochem-tools/pnautilus}.

\begin{acknowledgements}
The authors thank the anonymous referee for helpful comments.
T.-Y. Tu acknowledges the financial support of the China Scholarship Council (No. 202406190190). 
V.W. acknowledges the CNRS program ``Physique et Chimie du Milieu Interstellaire'' (PCMI) co-funded by the Centre National d’Etudes Spatiales (CNES).
Y.C. acknowledges the support from NSFC under grants Nos. 12573047, 12173018 and 12121003. 

\end{acknowledgements}

%
%

\bibliographystyle{aa}
\bibliography{2025_article_radiolysis}

@article{Aikawa_Grain_1999,
  title = {Grain {{Surface Recombination}} of {{HCO}}+},
  author = {Aikawa, Yuri and Herbst, Eric and Dzegilenko, Fedor N.},
  year = 1999,
  journal = {ApJ},
  volume = {527},
  pages = {262--265},
  publisher = {IOP},
  doi = {10.1086/308079}
}

@article{Atkinson_Evaluated_2004,
  title = {Evaluated Kinetic and Photochemical Data for Atmospheric Chemistry: {{Volume I}} - Gas Phase Reactions of {{Ox}}, {{HOx}}, {{NOx}} and {{SOx}} Species},
  shorttitle = {Evaluated Kinetic and Photochemical Data for Atmospheric Chemistry},
  author = {Atkinson, R. and Baulch, D. L. and Cox, R. A. and Crowley, J. N. and Hampson, R. F. and Hynes, R. G. and Jenkin, M. E. and Rossi, M. J. and Troe, J.},
  year = 2004,
  journal = {ACP},
  volume = {4},
  pages = {1461--1738},
  doi = {10.5194/acp-4-1461-200410.5194/acpd-3-6179-2003}
}

@article{Bacmann_Detection_2012,
  title = {Detection of Complex Organic Molecules in a Prestellar Core: A New Challenge for Astrochemical Models},
  shorttitle = {Detection of Complex Organic Molecules in a Prestellar Core},
  author = {Bacmann, A. and Taquet, V. and Faure, A. and Kahane, C. and Ceccarelli, C.},
  year = 2012,
  journal = {A\&A},
  volume = {541},
  pages = {L12},
  doi = {10.1051/0004-6361/201219207},
  langid = {english}
}

@article{Baragiola_Radiation_2008,
  title = {Radiation Effects in Ice: {{New}} Results},
  shorttitle = {Radiation Effects in Ice},
  author = {Baragiola, R. A. and Fam{\'a}, M. and Loeffler, M. J. and Raut, U. and Shi, J.},
  year = 2008,
  journal = {NIMPB},
  series = {Radiation {{Effects}} in {{Insulators}}},
  volume = {266},
  number = {12},
  pages = {3057--3062},
  doi = {10.1016/j.nimb.2008.03.186}
}

@article{Baragiola_Solid-state_1999,
  title = {Solid-State Ozone Synthesis by Energetic Ions},
  author = {Baragiola, R. A. and Atteberry, C. L. and Bahr, D. A. and Jakas, M. M.},
  year = 1999,
  journal = {NIMPB},
  volume = {157},
  pages = {233--238},
  doi = {10.1016/S0168-583X(99)00431-0}
}

@article{Baratta_comparison_2002,
  title = {A Comparison of Ion Irradiation and {{UV}} Photolysis of {{CH4}} and {{CH3OH}}},
  author = {Baratta, G. A. and Leto, G. and Palumbo, M. E.},
  year = 2002,
  journal = {A\&A},
  volume = {384},
  pages = {343--349},
  doi = {10.1051/0004-6361:20011835}
}

@article{Benderskii_Diffusion-limited_1996,
  title = {Diffusion-Limited Geminate Recombination of {{O}}+{{O2}} in Solid Xenon},
  author = {Benderskii, Alexander V. and Wight, Charles A.},
  year = 1996,
  journal = {JChPh},
  volume = {104},
  pages = {85--94},
  publisher = {AIP},
  doi = {10.1063/1.470877}
}

@article{Bennett_Laboratory_2005a,
  title = {Laboratory {{Studies}} on the {{Formation}} of {{Three C2H4O Isomers-Acetaldehyde}} ({{CH3CHO}}), {{Ethylene Oxide}} (c-{{C2H4O}}), and {{Vinyl Alcohol}} ({{CH2CHOH}})-in {{Interstellar}} and {{Cometary Ices}}},
  author = {Bennett, Chris J. and Osamura, Yoshihiro and Lebar, Matt D. and Kaiser, Ralf I.},
  year = 2005,
  journal = {ApJ},
  volume = {634},
  pages = {698--711},
  doi = {10.1086/452618}
}

@misc{Bialy_first_2025,
  title = {The First Detection of Cosmic-Ray Excited {{H}}\$\_2\$ in Interstellar Space},
  author = {Bialy, Shmuel and Chemke, Amit and Neufeld, David A. and Page, James Muzerolle and Ivlev, Alexei V. and Belli, Sirio and Gaches, Brandt A. L. and Godard, Benjamin and Bisbas, Thomas G. and Caselli, Paola and Jacob, Arshia M. and Padovani, Marco and Rab, Christian and Silsbee, Kedron and Porter, Troy A.},
  year = 2025,
  number = {arXiv:2508.20168},
  eprint = {2508.20168},
  primaryclass = {astro-ph},
  publisher = {arXiv},
  doi = {10.48550/arXiv.2508.20168},
  archiveprefix = {arXiv}
}

@article{Bisbas_Cosmic-ray_2017,
  title = {Cosmic-Ray {{Induced Destruction}} of {{CO}} in {{Star-forming Galaxies}}},
  author = {Bisbas, Thomas G. and {van Dishoeck}, Ewine F. and Papadopoulos, Padelis P. and Sz{\H u}cs, L{\'a}szl{\'o} and Bialy, Shmuel and Zhang, Zhi-Yu},
  year = 2017,
  journal = {ApJ},
  volume = {839},
  pages = {90},
  doi = {10.3847/1538-4357/aa696d}
}

@article{Bisbas_Effective_2015,
  title = {Effective {{Destruction}} of {{CO}} by {{Cosmic Rays}}: {{Implications}} for {{Tracing H2 Gas}} in the {{Universe}}},
  shorttitle = {Effective {{Destruction}} of {{CO}} by {{Cosmic Rays}}},
  author = {Bisbas, Thomas G. and Papadopoulos, Padelis P. and Viti, Serena},
  year = 2015,
  journal = {ApJ},
  volume = {803},
  pages = {37},
  publisher = {IOP},
  doi = {10.1088/0004-637X/803/1/37}
}

@article{Boduch_Chemistry_2012,
  title = {Chemistry Induced by Energetic Ions in Water Ice Mixed with Molecular Nitrogen and Oxygen},
  author = {Boduch, {\relax Ph}. and Domaracka, A. and Fulvio, D. and Langlinay, T. and Lv, X. Y. and Palumbo, M. E. and Rothard, H. and Strazzulla, G.},
  year = 2012,
  journal = {A\&A},
  volume = {544},
  pages = {A30},
  doi = {10.1051/0004-6361/201219365}
}

@article{Bovino_new_2020,
  title = {A New Proxy to Estimate the Cosmic Ray Ionization Rate in Dense Cores},
  author = {Bovino, S. and {Ferrada-Chamorro}, S. and Lupi, A. and Schleicher, D. R. G. and Caselli, P.},
  year = 2020,
  journal = {MNRAS},
  volume = {495},
  pages = {L7-L11},
  doi = {10.1093/mnrasl/slaa048}
}

@article{Bulak_Quantification_2022,
  title = {Quantification of {{O2}} Formation during {{UV}} Photolysis of Water Ice: {{H2O}} and {{H2O}}:{{CO2}} Ices},
  shorttitle = {Quantification of {{O2}} Formation during {{UV}} Photolysis of Water Ice},
  author = {Bulak, M. and Paardekooper, D. M. and Fedoseev, G. and Chuang, K. -J. and {Terwisscha van Scheltinga}, J. and Eistrup, C. and Linnartz, H.},
  year = 2022,
  journal = {A\&A},
  volume = {657},
  pages = {A120},
  doi = {10.1051/0004-6361/202141875}
}

@article{Buszek_Effects_2012,
  title = {Effects of a {{Single Water Molecule}} on the {{OH}} + {{H2O2Reaction}}},
  author = {Buszek, Robert J. and {Torrent-Sucarrat}, Miquel and Anglada, Josep M. and Francisco, Joseph S.},
  year = 2012,
  journal = {JPCA},
  volume = {116},
  pages = {5821--5829},
  doi = {10.1021/jp2077825}
}

@article{Caselli_Ionization_1998,
  title = {The {{Ionization Fraction}} in {{Dense Cloud Cores}}},
  author = {Caselli, P. and Walmsley, C. M. and Terzieva, R. and Herbst, Eric},
  year = 1998,
  journal = {ApJ},
  volume = {499},
  pages = {234--249},
  doi = {10.1086/305624}
}

@article{Chabot_Cosmic-ray_2016,
  title = {Cosmic-Ray Slowing down in Molecular Clouds: {{Effects}} of Heavy Nuclei},
  shorttitle = {Cosmic-Ray Slowing down in Molecular Clouds},
  author = {Chabot, Marin},
  year = 2016,
  journal = {A\&A},
  volume = {585},
  pages = {A15},
  doi = {10.1051/0004-6361/201425441}
}

@article{Chang_Gas-grain_2007,
  title = {Gas-Grain Chemistry in Cold Interstellar Cloud Cores with a Microscopic {{Monte Carlo}} Approach to Surface Chemistry},
  author = {Chang, Q. and Cuppen, H. M. and Herbst, E.},
  year = 2007,
  journal = {A\&A},
  volume = {469},
  pages = {973--983},
  doi = {10.1051/0004-6361:20077423}
}

@article{Cui_Exploring_2024,
  title = {Exploring the {{Role}} of {{Ion-Molecule Reactions}} on {{Interstellar Icy Grain Surfaces}}},
  author = {Cui, Weikai and Herbst, Eric},
  year = 2024,
  journal = {ACS Earth and Space Chemistry},
  volume = {8},
  pages = {2218--2231},
  publisher = {ACS},
  doi = {10.1021/acsearthspacechem.4c00194}
}

@article{Cuppen_Water_2010,
  title = {Water Formation at Low Temperatures by Surface {{O2}} Hydrogenation {{II}}: The Reaction Network},
  shorttitle = {Water Formation at Low Temperatures by Surface {{O2}} Hydrogenation {{II}}},
  author = {Cuppen, H. M. and Ioppolo, S. and Romanzin, C. and Linnartz, H.},
  year = 2010,
  journal = {PCCP},
  volume = {12},
  pages = {12077},
  doi = {10.1039/C0CP00251H}
}

@article{Dalgarno_Interstellar_2006,
  title = {Interstellar {{Chemistry Special Feature}}: {{The}} Galactic Cosmic Ray Ionization Rate},
  shorttitle = {Interstellar {{Chemistry Special Feature}}},
  author = {Dalgarno, A.},
  year = 2006,
  journal = {PNAS},
  volume = {103},
  pages = {12269--12273},
  doi = {10.1073/pnas.0602117103}
}

@article{Dartois_Cosmic_2018,
  title = {Cosmic Ray Sputtering Yield of Interstellar {{H2O}} Ice Mantles. {{Ice}} Mantle Thickness Dependence},
  author = {Dartois, E. and Chabot, M. and Id Barkach, T. and Rothard, H. and Aug{\'e}, B. and Agnihotri, A. N. and Domaracka, A. and Boduch, P.},
  year = 2018,
  journal = {A\&A},
  volume = {618},
  pages = {A173},
  doi = {10.1051/0004-6361/201833277}
}

@article{deBarros_Radiolysis_2011,
  title = {Radiolysis of Frozen Methanol by Heavy Cosmic Ray and Energetic Solar Particle Analogues},
  author = {{de Barros}, A. L. F. and Domaracka, A. and Andrade, D. P. P. and Boduch, P. and Rothard, H. and {da Silveira}, E. F.},
  year = 2011,
  journal = {MNRAS},
  volume = {418},
  pages = {1363--1374},
  publisher = {OUP},
  doi = {10.1111/j.1365-2966.2011.19587.x}
}

@article{Dewhurst_General_1952,
  title = {General Discussion},
  author = {Dewhurst, H. A. and Dale, W. M. and Bacq, Z. M. and Swallow, A. J. and Weiss, J. and Magat, M. and Minder, W. and Suttle, J. F. and Schulte, J. W. and Burton, Milton and Miller, N.},
  year = 1952,
  journal = {Discuss. Faraday Soc.},
  volume = {12},
  number = {0},
  pages = {312--318},
  publisher = {The Royal Society of Chemistry},
  doi = {10.1039/DF9521200312},
  langid = {english}
}

@article{Ellingson_Reactions_2007,
  title = {Reactions of {{Hydrogen Atom}} with {{Hydrogen Peroxide}}},
  author = {Ellingson, Benjamin A. and Theis, Daniel P. and Tishchenko, Oksana and Zheng, Jingjing and Truhlar, Donald G.},
  year = 2007,
  journal = {JCPA},
  volume = {111},
  pages = {13554--13566},
  doi = {10.1021/jp077379x}
}

@article{Ennis_formation_2011,
  title = {On the Formation of Ozone in Oxygen-Rich Solar System Ices via Ionizing Radiation},
  author = {Ennis, Courtney P. and Bennett, Chris J. and Kaiser, Ralf I.},
  year = 2011,
  journal = {PCCP},
  volume = {13},
  pages = {9469},
  doi = {10.1039/C1CP20434C}
}

@misc{Gaches_High-energy_2025,
  title = {High-Energy Astrochemistry in the Molecular Interstellar Medium},
  author = {Gaches, Brandt A. L. and Viti, Serena},
  year = 2025,
  number = {arXiv:2512.10060},
  eprint = {2512.10060},
  primaryclass = {astro-ph},
  publisher = {arXiv},
  doi = {10.48550/arXiv.2512.10060},
  archiveprefix = {arXiv}
}

@article{Garrod_Formation_2011,
  title = {On the {{Formation}} of {{CO2}} and {{Other Interstellar Ices}}},
  author = {Garrod, R. T. and Pauly, T.},
  year = 2011,
  journal = {ApJ},
  volume = {735},
  pages = {15},
  publisher = {IOP},
  doi = {10.1088/0004-637X/735/1/15}
}

@article{Garrod_new_2008,
  title = {A New Modified-Rate Approach for Gas-Grain Chemical Simulations},
  author = {Garrod, R. T.},
  year = 2008,
  journal = {A\&A},
  volume = {491},
  pages = {239--251},
  doi = {10.1051/0004-6361:200810518}
}

@article{Goldsmith_Molecular_1978,
  title = {Molecular Cooling and Thermal Balance of Dense Interstellar Clouds.},
  author = {Goldsmith, P. F. and Langer, W. D.},
  year = 1978,
  journal = {ApJ},
  volume = {222},
  pages = {881--895},
  publisher = {IOP},
  doi = {10.1086/156206}
}

@article{Gomis_Hydrogen_2004,
  title = {Hydrogen Peroxide Production by Ion Irradiation of Thin Water Ice Films},
  author = {Gomis, O. and Leto, G. and Strazzulla, G.},
  year = 2004,
  journal = {A\&A},
  volume = {420},
  pages = {405--410},
  doi = {10.1051/0004-6361:20041091}
}

@article{Gomis_Hydrogen_2004a,
  title = {Hydrogen Peroxide Formation by Ion Implantation in Water Ice and Its Relevance to the {{Galilean}} Satellites},
  author = {Gomis, O. and Satorre, M. A. and Strazzulla, G. and Leto, G.},
  year = 2004,
  journal = {P\&SS},
  volume = {52},
  pages = {371--378},
  publisher = {Elsevier},
  doi = {10.1016/j.pss.2003.06.010}
}

@article{Hasegawa_Models_1992,
  title = {Models of {{Gas-Grain Chemistry}} in {{Dense Interstellar Clouds}} with {{Complex Organic Molecules}}},
  author = {Hasegawa, Tatsuhiko I. and Herbst, Eric and Leung, Chun M.},
  year = 1992,
  journal = {ApJS},
  volume = {82},
  pages = {167},
  publisher = {IOP},
  doi = {10.1086/191713}
}

@article{Hasegawa_New_1993,
  title = {New Gas-Grain Chemical Models of Quiscent Dense Interstellar Clouds :The Effects of {{H2}} Tunnelling Reactions and Cosmic Ray Induced Desorption.},
  shorttitle = {New Gas-Grain Chemical Models of Quiscent Dense Interstellar Clouds},
  author = {Hasegawa, Tatsuhiko I. and Herbst, Eric},
  year = 1993,
  journal = {MNRAS},
  volume = {261},
  pages = {83--102},
  doi = {10.1093/mnras/261.1.83}
}

@article{Horl_Structure_1982,
  title = {Structure of Solid {$\alpha$}'-Oxygen},
  author = {H{\"o}rl, E. M. and Kohlbeck, F.},
  year = 1982,
  journal = {Acta Crystallographica Section B: Structural Crystallography and Crystal Chemistry},
  volume = {38},
  number = {1},
  pages = {20--23},
  publisher = {International Union of Crystallography},
  doi = {10.1107/S0567740882001952},
  langid = {english}
}

@article{Hudson_Laboratory_1999,
  title = {Laboratory {{Studies}} of the {{Formation}} of {{Methanol}} and {{Other Organic Molecules}} by {{Water}}+{{Carbon Monoxide Radiolysis}}: {{Relevance}} to {{Comets}}, {{Icy Satellites}}, and {{Interstellar Ices}}},
  shorttitle = {Laboratory {{Studies}} of the {{Formation}} of {{Methanol}} and {{Other Organic Molecules}} by {{Water}}+{{Carbon Monoxide Radiolysis}}},
  author = {Hudson, R. L. and Moore, M. H.},
  year = 1999,
  journal = {Icarus},
  volume = {140},
  pages = {451--461},
  doi = {10.1006/icar.1999.6144}
}

@article{Indriolo_Cosmic-ray_2013,
  title = {Cosmic-Ray Astrochemistry},
  author = {Indriolo, Nick and McCall, Benjamin J.},
  year = 2013,
  journal = {ChSRv},
  volume = {42},
  pages = {7763--7773},
  doi = {10.1039/C3CS60087D}
}

@article{Indriolo_Investigating_2012,
  title = {Investigating the {{Cosmic-Ray Ionization Rate}} in the {{Galactic Diffuse Interstellar Medium}} through {{Observations}} of {{H}}+ 3},
  author = {Indriolo, Nick and McCall, Benjamin J.},
  year = 2012,
  journal = {ApJ},
  volume = {745},
  pages = {91},
  publisher = {IOP},
  doi = {10.1088/0004-637X/745/1/91}
}

@article{Indriolo_Mapping_2026,
  title = {Mapping the {{Cosmic-Ray Ionization Rate}} in the {{Local Galaxy}} with {{H3}}+},
  author = {Indriolo, Nick and Ivlev, Alexei V. and Pellegrin, T. and Obolentseva, M. and Caselli, Paola and Jacob, A. M. and Neufeld, D. A. and Silsbee, Kedron and Wolfire, M. G.},
  year = 2026,
  journal = {ApJ},
  volume = {997},
  pages = {123},
  publisher = {IOP},
  doi = {10.3847/1538-4357/ae21ba}
}

@article{Ivlev_Impulsive_2015,
  title = {Impulsive {{Spot Heating}} and {{Thermal Explosion}} of {{Interstellar Grains Revisited}}},
  author = {Ivlev, A. V. and R{\"o}cker, T. B. and Vasyunin, A. and Caselli, P.},
  year = 2015,
  journal = {ApJ},
  volume = {805},
  pages = {59},
  doi = {10.1088/0004-637X/805/1/59}
}

@article{Jin_Formation_2020,
  title = {Formation of {{Complex Organic Molecules}} in {{Cold Interstellar Environments}} through {{Nondiffusive Grain-surface}} and {{Ice-mantle Chemistry}}},
  author = {Jin, Mihwa and Garrod, Robin T.},
  year = 2020,
  journal = {ApJS},
  volume = {249},
  pages = {26},
  publisher = {IOP},
  doi = {10.3847/1538-4365/ab9ec8}
}

@article{Kalvans_Temperature_2018,
  title = {Temperature {{Spectra}} of {{Interstellar Dust Grains Heated}} by {{Cosmic Rays}}. {{II}}. {{Dark Cloud Cores}}},
  author = {Kalv{\=a}ns, Juris},
  year = 2018,
  journal = {ApJS},
  volume = {239},
  pages = {6},
  publisher = {IOP},
  doi = {10.3847/1538-4365/aae527}
}

@article{Lamberts_Influence_2017,
  title = {Influence of {{Surface}} and {{Bulk Water Ice}} on the {{Reactivity}} of a {{Water-forming Reaction}}},
  author = {Lamberts, Thanja and K{\"a}stner, Johannes},
  year = 2017,
  journal = {ApJ},
  volume = {846},
  pages = {43},
  publisher = {IOP},
  doi = {10.3847/1538-4357/aa8311}
}

@article{Lamberts_Quantum_2016,
  title = {Quantum Tunneling during Interstellar Surface-Catalyzed Formation of Water: The Reaction {{H}} + {{H2O2}} {$\rightarrow$} {{H2O}} + {{OH}}},
  shorttitle = {Quantum Tunneling during Interstellar Surface-Catalyzed Formation of Water},
  author = {Lamberts, Thanja and Samanta, Pradipta Kumar and K{\"o}hn, Andreas and K{\"a}stner, Johannes},
  year = 2016,
  journal = {PCCP},
  volume = {18},
  pages = {33021--33030},
  doi = {10.1039/C6CP06457D}
}

@article{Lamberts_Water_2013,
  title = {Water Formation at Low Temperatures by Surface {{O2}} Hydrogenation {{III}}: {{Monte Carlo}} Simulation},
  shorttitle = {Water Formation at Low Temperatures by Surface {{O2}} Hydrogenation {{III}}},
  author = {Lamberts, Thanja and Cuppen, Herma M. and Ioppolo, Sergio and Linnartz, Harold},
  year = 2013,
  journal = {PCCP},
  volume = {15},
  pages = {8287},
  doi = {10.1039/C3CP00106G}
}

@article{LeCaer_Water_2011,
  title = {Water {{Radiolysis}}: {{Influence}} of {{Oxide Surfaces}} on {{H2 Production}} under {{Ionizing Radiation}}},
  shorttitle = {Water {{Radiolysis}}},
  author = {Le Ca{\"e}r, Sophie},
  year = 2011,
  journal = {Water},
  volume = {3},
  number = {1},
  pages = {235--253},
  publisher = {Molecular Diversity Preservation International},
  doi = {10.3390/w3010235},
  copyright = {http://creativecommons.org/licenses/by/3.0/},
  langid = {english}
}

@article{Leger_Desorption_1985,
  title = {Desorption from Interstellar Grains},
  author = {L{\'e}ger, A. and Jura, M. and Omont, A.},
  year = 1985,
  journal = {A\&A},
  volume = {144},
  pages = {147--160}
}

@article{Loeffler_Synthesis_2006,
  title = {Synthesis of Hydrogen Peroxide in Water Ice by Ion Irradiation},
  author = {Loeffler, M. J. and Raut, U. and Vidal, R. A. and Baragiola, R. A. and Carlson, R. W.},
  year = 2006,
  journal = {Icarus},
  volume = {180},
  pages = {265--273},
  doi = {10.1016/j.icarus.2005.08.001}
}

@article{Lu_Rate_2018,
  title = {Rate Coefficients of the {{H}} + {{H2O2}} {$\rightarrow$} {{H2}} + {{HO2}} Reaction on an Accurate Fundamental Invariant-Neural Network Potential Energy Surface},
  author = {Lu, Xiaoxiao and Meng, Qingyong and Wang, Xingan and Fu, Bina and Zhang, Dong H.},
  year = 2018,
  journal = {JChPh},
  volume = {149},
  pages = {174303},
  publisher = {AIP},
  doi = {10.1063/1.5063613}
}

@article{Lu_Theoretical_2019,
  title = {Theoretical {{Investigations}} of {{Rate Coefficients}} of {{H}} + {{H2O2}} {$\rightarrow$} {{OH}} + {{H2O}} on a {{Full-Dimensional Potential Energy Surface}}},
  author = {Lu, Xiaoxiao and Wang, Xingan and Fu, Bina and Zhang, Donghui},
  year = 2019,
  journal = {JPCA},
  volume = {123},
  number = {18},
  pages = {3969--3976},
  publisher = {American Chemical Society},
  doi = {10.1021/acs.jpca.9b02526}
}

@article{Mejia_Swift_2022,
  title = {Swift Heavy Ions Irradiation of Water Ice at Different Temperatures: Hydrogen Peroxide and Ozone Synthesis and Sputtering Yield},
  shorttitle = {Swift Heavy Ions Irradiation of Water Ice at Different Temperatures},
  author = {Mej{\'i}a, C. and {de Barros}, A. L. F. and Rothard, H. and Boduch, P. and {da Silveira}, E. F.},
  year = 2022,
  journal = {MNRAS},
  volume = {514},
  pages = {3789--3801},
  publisher = {OUP},
  doi = {10.1093/mnras/stac1489}
}

@article{Minissale_CO2_2013,
  title = {{{CO2}} Formation on Interstellar Dust Grains: A Detailed Study of the Barrier of the {{CO}} + {{O}} Channel},
  shorttitle = {{{CO2}} Formation on Interstellar Dust Grains},
  author = {Minissale, M. and Congiu, E. and Manic{\`o}, G. and Pirronello, V. and Dulieu, F.},
  year = 2013,
  journal = {A\&A},
  volume = {559},
  pages = {A49},
  publisher = {EDP},
  doi = {10.1051/0004-6361/201321453}
}

@article{Moore_IR_2000,
  title = {{{IR Detection}} of {{H 2O}} 2 at 80 {{K}} in {{Ion-Irradiated Laboratory Ices Relevant}} to {{Europa}}},
  author = {Moore, M. H. and Hudson, R. L.},
  year = 2000,
  journal = {Icarus},
  volume = {145},
  pages = {282--288},
  publisher = {Elsevier},
  doi = {10.1006/icar.1999.6325}
}

@article{Mullikin_New_2021,
  title = {A {{New Method}} for {{Simulating Photoprocesses}} in {{Astrochemical Models}}},
  author = {Mullikin, Ella and Anderson, Hannah and O'Hern, Natalie and Farrah, Megan and Arumainayagam, Christopher R. and {van Dishoeck}, Ewine F. and Gerakines, Perry A. and Vasyunin, Anton I. and Majumdar, Liton and Caselli, Paola and Shingledecker, Christopher N.},
  year = 2021,
  journal = {ApJ},
  volume = {910},
  pages = {72},
  doi = {10.3847/1538-4357/abd778}
}

@article{Muroya_re-evaluation_2005,
  title = {A Re-Evaluation of the Initial Yield of the Hydrated Electron in the Picosecond Time Range},
  author = {Muroya, Yusa and Lin, Mingzhang and Wu, Guozhong and Iijima, Hokuto and Yoshii, Koji and Ueda, Toru and Kudo, Hisaaki and Katsumura, Yosuke},
  year = 2005,
  journal = {RaPC},
  volume = {72},
  pages = {169--172},
  publisher = {Elsevier},
  doi = {10.1016/j.radphyschem.2004.09.011}
}

@article{Neufeld_Cosmic-Ray_2017,
  title = {The {{Cosmic-Ray Ionization Rate}} in the {{Galactic Disk}}, as {{Determined}} from {{Observations}} of {{Molecular Ions}}},
  author = {Neufeld, David A. and Wolfire, Mark G.},
  year = 2017,
  journal = {ApJ},
  volume = {845},
  pages = {163},
  doi = {10.3847/1538-4357/aa6d68}
}

@article{Neufeld_Densities_2024a,
  title = {The {{Densities}} in {{Diffuse}} and {{Translucent Molecular Clouds}}: {{Estimates}} from {{Observations}} of {{C2}} and from {{Three-dimensional Extinction Maps}}},
  shorttitle = {The {{Densities}} in {{Diffuse}} and {{Translucent Molecular Clouds}}},
  author = {Neufeld, David A. and Welty, Daniel E. and Ivlev, Alexei V. and Caselli, Paola and Edenhofer, Gordian and Indriolo, Nick and Obolentseva, Marta and Silsbee, Kedron and Sonnentrucker, Paule and Wolfire, Mark G.},
  year = 2024,
  journal = {ApJ},
  volume = {973},
  pages = {143},
  publisher = {IOP},
  doi = {10.3847/1538-4357/ad7264}
}

@misc{Neufeld_JWST_2025,
  title = {{{JWST}} Observations of Cosmic-Ray-Excited {{H}}\$\_2\$ in {{Barnard}} 68: Spatial Variations and Constraints on Cosmic-Ray Attenuation},
  shorttitle = {{{JWST}} Observations of Cosmic-Ray-Excited {{H}}\$\_2\$ in {{Barnard}} 68},
  author = {Neufeld, David A. and Silsbee, Kedron and Ivlev, Alexei V. and Bialy, Shmuel and Gaches, Brandt A. L. and Padovani, Marco and Belli, Sirio and Bisbas, Thomas G. and Chemke, Amit and Godard, Benjamin and Page, James Muzerolle and Rab, Christian},
  year = 2025,
  number = {arXiv:2511.16003},
  eprint = {2511.16003},
  primaryclass = {astro-ph},
  publisher = {arXiv},
  doi = {10.48550/arXiv.2511.16003},
  archiveprefix = {arXiv}
}

@article{Oberg_Photodesorption_2007,
  title = {Photodesorption of {{CO Ice}}},
  author = {{\"O}berg, Karin I. and Fuchs, Guido W. and Awad, Zainab and Fraser, Helen J. and Schlemmer, Stephan and {van Dishoeck}, Ewine F. and Linnartz, Harold},
  year = 2007,
  journal = {ApJ},
  volume = {662},
  pages = {L23-L26},
  publisher = {IOP},
  doi = {10.1086/519281}
}

@article{Obolentseva_Reevaluation_2024,
  title = {Reevaluation of the {{Cosmic-Ray Ionization Rate}} in {{Diffuse Clouds}}},
  author = {Obolentseva, M. and Ivlev, A. V. and Silsbee, K. and Neufeld, D. A. and Caselli, P. and Edenhofer, G. and Indriolo, N. and Bisbas, T. G. and Lomeli, D.},
  year = 2024,
  journal = {ApJ},
  volume = {973},
  pages = {142},
  publisher = {IOP},
  doi = {10.3847/1538-4357/ad71ce}
}

@article{Padovani_Cosmic-ray_2009,
  title = {Cosmic-Ray Ionization of Molecular Clouds},
  author = {Padovani, M. and Galli, D. and Glassgold, A. E.},
  year = 2009,
  journal = {A\&A},
  volume = {501},
  number = {2},
  pages = {619--631},
  doi = {10.1051/0004-6361/200911794},
  langid = {english}
}

@article{Palumbo_ROCN_2000,
  title = {{{ROCN Species Produced}} by {{Ion Irradiation}} of {{Ice Mixtures}}: {{Comparison}} with {{Astronomical Observations}}},
  shorttitle = {{{ROCN Species Produced}} by {{Ion Irradiation}} of {{Ice Mixtures}}},
  author = {Palumbo, M. E. and Strazzulla, G. and Pendleton, Y. J. and Tielens, A. G. G. M.},
  year = 2000,
  journal = {ApJ},
  volume = {534},
  pages = {801--808},
  publisher = {IOP},
  doi = {10.1086/308760}
}

@article{Paulive_role_2021,
  title = {The Role of Radiolysis in the Modelling of {{C2H4O2}} Isomers and Dimethyl Ether in Cold Dark Clouds},
  author = {Paulive, Alec and Shingledecker, Christopher N. and Herbst, Eric},
  year = 2021,
  journal = {MNRAS},
  volume = {500},
  pages = {3414--3424},
  doi = {10.1093/mnras/staa3458}
}

@article{Pilling_Chemical_2022,
  title = {Chemical {{Evolution}} of {{CO2 Ices}} under {{Processing}} by {{Ionizing Radiation}}: {{Characterization}} of {{Nonobserved Species}} and {{Chemical Equilibrium Phase}} with the {{Employment}} of {{PROCODA Code}}},
  shorttitle = {Chemical {{Evolution}} of {{CO2 Ices}} under {{Processing}} by {{Ionizing Radiation}}},
  author = {Pilling, Sergio and Carvalho, Geanderson A. and Rocha, Will R. M.},
  year = 2022,
  journal = {ApJ},
  volume = {925},
  pages = {147},
  doi = {10.3847/1538-4357/ac3d8a}
}

@article{Pilling_Chemical_2023,
  title = {Chemical Evolution of Electron-Bombarded Crystalline Water Ices at Different Temperatures Using the {{PROCODA}} Code},
  author = {Pilling, S. and {da Silveira}, C. H. and {Ojeda-Gonzalez}, A.},
  year = 2023,
  journal = {MNRAS},
  volume = {523},
  pages = {2858--2875},
  publisher = {OUP},
  doi = {10.1093/mnras/stad1518}
}

@article{Pilling_Radiolysis_2010,
  title = {Radiolysis of Ammonia-Containing Ices by Energetic, Heavy, and Highly Charged Ions inside Dense Astrophysical Environments},
  author = {Pilling, S. and Seperuelo Duarte, E. and {da Silveira}, E. F. and Balanzat, E. and Rothard, H. and Domaracka, A. and Boduch, P.},
  year = 2010,
  journal = {A\&A},
  volume = {509},
  pages = {A87},
  doi = {10.1051/0004-6361/200912274}
}

@article{Prasad_UV_1983,
  title = {{{UV}} Radiation Field inside Dense Clouds - {{Its}} Possible Existence and Chemical Implications},
  author = {Prasad, S. S. and Tarafdar, S. P.},
  year = 1983,
  journal = {ApJ},
  volume = {267},
  pages = {603--609},
  publisher = {IOP},
  doi = {10.1086/160896}
}

@article{Ruaud_Gas_2016,
  title = {Gas and Grain Chemical Composition in Cold Cores as Predicted by the {{Nautilus}} Three-Phase Model},
  author = {Ruaud, Maxime and Wakelam, Valentine and Hersant, Franck},
  year = 2016,
  journal = {MNRAS},
  volume = {459},
  pages = {3756--3767},
  doi = {10.1093/mnras/stw887}
}

@article{Rudd_Cross_1983,
  title = {Cross Sections for Ionization of Gases by 5-4000-{{keV}} Protons and for Electron Capture by 5-150-{{keV}} Protons},
  author = {Rudd, M. E. and Goffe, T. V. and Dubois, R. D. and Toburen, L. H. and Ratcliffe, C. A.},
  year = 1983,
  journal = {PhRvA},
  volume = {28},
  pages = {3244--3257},
  publisher = {APS},
  doi = {10.1103/PhysRevA.28.3244}
}

@article{SeperueloDuarte_Laboratory_2010,
  title = {Laboratory Simulation of Heavy-Ion Cosmic-Ray Interaction with Condensed {{CO}}},
  author = {Seperuelo Duarte, E. and Domaracka, A. and Boduch, P. and Rothard, H. and Dartois, E. and {da Silveira}, E. F.},
  year = 2010,
  journal = {A\&A},
  volume = {512},
  pages = {A71},
  doi = {10.1051/0004-6361/200912899}
}

@article{Shen_Cosmic_2004,
  title = {Cosmic Ray Induced Explosive Chemical Desorption in Dense Clouds},
  author = {Shen, C. J. and Greenberg, J. M. and Schutte, W. A. and {van Dishoeck}, E. F.},
  year = 2004,
  journal = {A\&A},
  volume = {415},
  pages = {203--215},
  doi = {10.1051/0004-6361:20031669}
}

@article{Shingledecker_Cosmic-Ray-driven_2018,
  title = {On {{Cosmic-Ray-driven Grain Chemistry}} in {{Cold Core Models}}},
  author = {Shingledecker, Christopher N. and Tennis, Jessica and Le Gal, Romane and Herbst, Eric},
  year = 2018,
  journal = {ApJ},
  volume = {861},
  pages = {20},
  doi = {10.3847/1538-4357/aac5ee}
}

@article{Shingledecker_general_2018,
  title = {A General Method for the Inclusion of Radiation Chemistry in Astrochemical Models},
  author = {Shingledecker, Christopher N. and Herbst, Eric},
  year = 2018,
  journal = {PCCP},
  volume = {20},
  pages = {5359--5367},
  doi = {10.1039/C7CP05901A}
}

@article{Shingledecker_new_2017,
  title = {A New Model of the Chemistry of Ionizing Radiation in Solids: {{CIRIS}}},
  shorttitle = {A New Model of the Chemistry of Ionizing Radiation in Solids},
  author = {Shingledecker, Christopher N. and Le Gal, Romane and Herbst, Eric},
  year = 2017,
  journal = {PCCP},
  volume = {19},
  pages = {11043--11056},
  doi = {10.1039/C7CP01472D}
}

@article{Shingledecker_Simulating_2019,
  title = {On {{Simulating}} the {{Proton-irradiation}} of {{O2}} and {{H2O Ices Using Astrochemical-type Models}}, with {{Implications}} for {{Bulk Reactivity}}},
  author = {Shingledecker, Christopher N. and Vasyunin, Anton and Herbst, Eric and Caselli, Paola},
  year = 2019,
  journal = {ApJ},
  volume = {876},
  pages = {140},
  doi = {10.3847/1538-4357/ab16d5}
}

@article{Silva_Molecular_2025,
  title = {Molecular {{Evolution}} of {{H2O}}:{{O2 Ices}} at {{Different Temperatures}} in {{Simulated Space Environments}}. {{I}}. {{Chemical Kinetics}} and {{Equilibrium}}},
  shorttitle = {Molecular {{Evolution}} of {{H2O}}},
  author = {Silva, J. R. C. and Queiroz, L. M. S. V. and Ferr{\~a}o, L. F. A. and Pilling, S.},
  year = 2025,
  journal = {ApJ},
  volume = {985},
  pages = {254},
  publisher = {IOP},
  doi = {10.3847/1538-4357/adc924}
}

@article{Spitzer_Heating_1968,
  title = {Heating of {{H}} i {{Regions}} by {{Energetic Particles}}},
  author = {Spitzer, Jr., Lyman and Tomasko, Martin G.},
  year = 1968,
  journal = {ApJ},
  volume = {152},
  pages = {971},
  publisher = {IOP},
  doi = {10.1086/149610}
}

@article{Talbi_interstellar_2006,
  title = {The Interstellar Gas-Phase Formation of {{CO}} 2 - {{Assisted}} or Not by Water Molecules?},
  author = {Talbi, D. and Chandler, G. S. and Rohl, A. L.},
  year = 2006,
  journal = {CP},
  volume = {320},
  pages = {214--228},
  publisher = {Elsevier},
  doi = {10.1016/j.chemphys.2005.07.033}
}

@article{Teolis_Water_2017,
  title = {Water {{Ice Radiolytic O2}}, {{H2}}, and {{H2O2 Yields}} for {{Any Projectile Species}}, {{Energy}}, or {{Temperature}}: {{A Model}} for {{Icy Astrophysical Bodies}}},
  shorttitle = {Water {{Ice Radiolytic O2}}, {{H2}}, and {{H2O2 Yields}} for {{Any Projectile Species}}, {{Energy}}, or {{Temperature}}},
  author = {Teolis, B. D. and Plainaki, C. and Cassidy, T. A. and Raut, U.},
  year = 2017,
  journal = {JGRE},
  volume = {122},
  pages = {1996--2012},
  publisher = {Wiley},
  doi = {10.1002/2017JE005285}
}

@article{Umebayashi_Recombination_1980,
  title = {Recombination of {{Ions}} and {{Electrons}} on {{Grains}} and the {{Ionization Degree}} in {{Dense Interstellar Clouds}}},
  author = {Umebayashi, Toyoharu and Nakano, Takenori},
  year = 1980,
  journal = {PASJ},
  volume = {32},
  pages = {405--421},
  publisher = {OUP},
  doi = {10.1093/pasj/32.3.405}
}

@article{Vasyunin_Formation_2017,
  title = {Formation of {{Complex Molecules}} in {{Prestellar Cores}}: {{A Multilayer Approach}}},
  shorttitle = {Formation of {{Complex Molecules}} in {{Prestellar Cores}}},
  author = {Vasyunin, A. I. and Caselli, P. and Dulieu, F. and {Jim{\'e}nez-Serra}, I.},
  year = 2017,
  journal = {ApJ},
  volume = {842},
  pages = {33},
  publisher = {IOP},
  doi = {10.3847/1538-4357/aa72ec}
}

@article{Vasyunin_Unified_2013,
  title = {A {{Unified Monte Carlo Treatment}} of {{Gas-Grain Chemistry}} for {{Large Reaction Networks}}. {{II}}. {{A Multiphase Gas-surface-layered Bulk Model}}},
  author = {Vasyunin, A. I. and Herbst, Eric},
  year = 2013,
  journal = {ApJ},
  volume = {762},
  pages = {86},
  publisher = {IOP},
  doi = {10.1088/0004-637X/762/2/86}
}

@article{Wakelam_2024_2024a,
  title = {The 2024 {{KIDA}} Network for Interstellar Chemistry},
  author = {Wakelam, V. and Gratier, P. and Loison, J. -C. and Hickson, K. M. and Penguen, J. and Mechineau, A.},
  year = 2024,
  journal = {A\&A},
  volume = {689},
  pages = {A63},
  publisher = {EDP},
  doi = {10.1051/0004-6361/202450606}
}

@article{Wakelam_Binding_2017,
  title = {Binding Energies: {{New}} Values and Impact on the Efficiency of Chemical Desorption},
  shorttitle = {Binding Energies},
  author = {Wakelam, V. and Loison, J. -C. and Mereau, R. and Ruaud, M.},
  year = 2017,
  journal = {MolAs},
  volume = {6},
  pages = {22--35},
  doi = {10.1016/j.molap.2017.01.002}
}

@article{Wakelam_Efficiency_2021,
  title = {Efficiency of Non-Thermal Desorptions in Cold-Core Conditions. {{Testing}} the Sputtering of Grain Mantles Induced by Cosmic Rays},
  author = {Wakelam, V. and Dartois, E. and Chabot, M. and Spezzano, S. and {Navarro-Almaida}, D. and Loison, J.-C. and Fuente, A.},
  year = 2021,
  journal = {A\&A},
  volume = {652},
  pages = {A63},
  doi = {10.1051/0004-6361/202039855},
  langid = {english}
}

@article{Zhang_Water-catalyzed_2011,
  title = {Water-Catalyzed Gas-Phase Hydrogen Abstraction Reactions of {{CH3O2}} and {{HO2}} with {{HO2}}: A Computational Investigation},
  shorttitle = {Water-Catalyzed Gas-Phase Hydrogen Abstraction Reactions of {{CH3O2}} and {{HO2}} with {{HO2}}},
  author = {Zhang, Tianlei and Wang, Wenliang and Zhang, Pei and L{\"u}, Jian and Zhang, Yue},
  year = 2011,
  journal = {PCCP},
  volume = {13},
  pages = {20794},
  doi = {10.1039/C1CP21563A}
}

@article{Zheng_Temperature_2006,
  title = {Temperature {{Dependence}} of the {{Formation}} of {{Hydrogen}}, {{Oxygen}}, and {{Hydrogen Peroxide}} in {{Electron-Irradiated Crystalline Water Ice}}},
  author = {Zheng, Weijun and Jewitt, David and Kaiser, Ralf I.},
  year = 2006,
  journal = {ApJ},
  volume = {648},
  pages = {753--761},
  doi = {10.1086/505901}
}

@article{Ziegler_SRIM_2010,
  title = {{{SRIM}} - {{The}} Stopping and Range of Ions in Matter (2010)},
  author = {Ziegler, James F. and Ziegler, M. D. and Biersack, J. P.},
  year = 2010,
  journal = {NIMPB},
  volume = {268},
  pages = {1818--1823},
  doi = {10.1016/j.nimb.2010.02.091}
}




\end{document}